\documentclass[12pt]{article}
\usepackage{amssymb,epsfig}
\renewcommand{\theequation}{\arabic{section}.\arabic{equation}}
\renewcommand{\thefootnote}{\fnsymbol{footnote}}

\newcommand{\be}[1]{
\begin{equation}\label{#1}}
\newcommand{\ee}{\end{equation}}
\newcommand{\ba}[1]{
\begin{eqnarray}\label{#1}}
\newcommand{\ea}{\end{eqnarray}}
\newcommand{\baa}{\begin{eqnarray*}}
\newcommand{\btab}{\begin{tabular}}
\newcommand{\etab}{\end{tabular}}
\newcommand{\eaa}{\end{eqnarray*}}

\newcommand{\deriv}{{\stackrel{\leftrightarrow}{D}}\!}
\newcommand{\derleft}{{\stackrel{\leftarrow}{D}}\!}
\newcommand{\derright}{{\stackrel{\rightarrow}{D}}\!}

\def \labeltest #1 {\label{#1}}
\newcommand\re[1]{(\ref{#1})}

\def \qqqquad {\qquad\qquad}

\def\II{\hbox{{1}\kern-.25em\hbox{l}}}

\def \Tr {\mbox{Tr\,}}

\newcommand \vev [1] {\langle{#1}\rangle}

\def \e {\mbox{e}}

\begin{document}

\begin{titlepage}
\begin{flushright}
\begin{tabular}{l}
LPT--Orsay--00--82\\
SPbU--IP--00--15\\
TPR--00--18\\
hep-ph/0010128
\end{tabular}
\end{flushright}
\vskip1cm
\begin{center}
  {\large \bf
     Gluon contribution to the structure function $g_2(x,Q^2)$
  \\}
\vspace{1cm}
{\sc V.M.~Braun}${}^{1}$,
{\sc G.P.~Korchemsky}${}^2$
          and {\sc A.N.~Manashov}${}^{1,3}$
\\[0.5cm]
\vspace*{0.1cm} ${}^1${\it
   Institut f\"ur Theoretische Physik, Universit\"at
   Regensburg, \\ D-93040 Regensburg, Germany
                       } \\[0.2cm]
\vspace*{0.1cm} ${}^2$ {\it
Laboratoire de Physique Th\'eorique%
\def\thefootnote{\fnsymbol{footnote}}%
\footnote{Unite Mixte de Recherche du CNRS (UMR 8627)},
Universit\'e de Paris XI, \\
91405 Orsay C\'edex, France
                       } \\[0.2cm]
\vspace*{0.1cm} ${}^3$ {\it
Department of Theoretical Physics,  Sankt-Petersburg State
University, \\ St.-Petersburg, Russia
                       } \\[1.0cm]

\vskip0.8cm
{\bf Abstract:\\[10pt]} \parbox[t]{\textwidth}{
We calculate the one-loop twist-3
gluon contribution to the flavor-singlet
structure function $g_2(x,Q^2)$ in polarized deep-inelastic
scattering and find that
it is dominated by the contribution of the three-gluon operator
with the lowest anomalous dimension (for each moment $N$).
The similar property was observed earlier for the nonsinglet
distributions, although the reason is in our case different.
The result is encouraging and suggests a simple
evolution pattern of $g_2(x,Q^2)$ in analogy with the conventional
description of twist-2 parton distributions.
}
\vskip1cm

\end{center}

\end{titlepage}

\def\thefootnote{\arabic{footnote}}
\setcounter{footnote} 0

\section{Introduction}
\setcounter{equation}{0}
Twist-three parton distributions in the nucleon are attracting
increasing interest as unique probes of quark-gluon correlations
in hadrons. They have clear experimental signatures and give rise to various
spin asymmetries in experiments with polarized beams and targets.
Quantitative studies of such asymmetries are becoming possible
with the increasing precision of experimental data at SLAC and RHIC,
and can constitute an important part of the future spin physics program
on high-luminosity accelerators like ELFE, eRHIC, etc.

The structure function $g_2(x,Q^2)$ in polarized deep inelastic scattering
presents the classical example of a twist-3 observable and  received
considerable attention in the past. The  experimental studies at SLAC
\cite{E143,E154,E155} have confirmed theoretical expectations about the
shape of $g_2(x,Q^2)$ and provided first evidence on the
most interesting twist-3 contribution. On the theoretical side,
a lot of effort was invested to understand the physical interpretation
of twist-3 distributions (see e.g. \cite{Book,AELreview,KTreview} for the
review of various aspects) and
their scale dependence \cite{SV82,BKL84,BB89,ABH,Kodaira,Muller,BKM00}.
Conditions for the validity of the Burhardt-Cottingham (BC) sum rule
\cite{BC70} $\int_0^1\! dx\, g_2(x)=0$ were discussed in detail
\cite{Book,JJ91,KodairaBC,Nikolaev} and
the second moment $\int_0^1\! dx\, x^2 g_2(x)$ was estimated using QCD sum rules
\cite{BBK90} and on the lattice \cite{Gockeler}.

In spite of the  significant progress that has been achieved,
understanding of the scale dependence of $g_2(x,Q^2)$ still poses an
outstanding theoretical problem. To explain the difficulty, we remind
that to the tree-level accuracy the structure function $g_2(x,Q^2)$
or, equivalently, $g_T(x,Q^2) =  g_1(x,Q^2)+g_2(x,Q^2)$ is given by
the quark-antiquark light-cone correlation function
in a transversely polarized nucleon (see, e.g. \cite{JJ91})
\ba{gT}
 g_T(x) &=& \frac{1}{2M} \int_{-\infty}^\infty\frac{d\lambda}{2\pi}\,
 {\rm e}^{i\lambda x}
 \langle p,s_\perp|\bar q(0)\gamma_\perp\gamma_5 q(\lambda n)|p,s_\perp\rangle,
\ea
where $p_\mu$ and $s_\mu$ are nucleon momentum and spin vectors, respectively,
and we assumed that the nucleon is moving in the
$z-$direction,%
\footnote{Throughout the paper we shall use the following definition of the
light-cone components $p_\pm=(p_0\pm p_3)/\sqrt{2}$ and $p_\perp=(p_1,p_2)$.
In addition, we shall not display the gauge factors connecting the quark fields
and ensuring the gauge invariance of nonlocal light-cone operators.}
$p_\mu=(p_+,p_-,{\mathbf 0}_\perp)$, $p^2=2p_+p_-=M^2$ and $p\cdot s=0$. The
light-like vector $n$ is given by $n_\mu=(0_+,1/p_+,{\mathbf 0}_\perp)$ so that
$n^2=0$, $(pn)=1$ and the transverse direction is defined as orthogonal to the
$(p,n)$ plane.
For comparison, the leading twist-2 spin structure function is written as
\ba{g1}
 g_1(x) &=& \int_{-\infty}^\infty \frac{d\lambda}{2\pi}\,
 {\rm e}^{i\lambda x}
 \langle p,s_z|\bar q(0)\!\not{\!n} \gamma_5 q(\lambda n)|p,s_z\rangle\,.
\ea

In the parton model, $g_1(x)$ measures the quark helicity distribution
in a longitudinally polarized nucleon. Such an interpretation
can be made because the quark helicity operator $\Sigma_p$ commutes with
the free-quark Dirac Hamiltonian $H= \alpha_z p_z$. On the contrary,
the quark spin operator projected along the transverse direction
$\Sigma_\perp$ does not commute with the Hamiltonian and thus there exists
no energy eigenstate $|p_z\rangle$ such that $\Sigma_\perp |p_z\rangle
= s_\perp |p_z\rangle$. The transverse
spin of the nucleon cannot, therefore,  be thought of as being composed of
transverse spins of individual quark (gluon) constituents.
 The transverse spin {\em average} of quarks
 in (\ref{gT}) that defines $g_T(x)$ is sensitive to the dynamics of
quark-gluon interactions and does not have any probabilistic interpretation
in terms of one-particle quark parton densities.

One possible way to see the relation of the transverse spin to gluonic
degrees of freedom is
to decompose the quark field operator in ``good'' (+) and
``bad'' (--) components $q(x) = q_+(x) + q_-(x)$ where
$P_+ = \frac12 \gamma_-\gamma_+$ and $P_- = \frac12 \gamma_+\gamma_-$ are
the corresponding projection operators \cite{KS70}. It is easy to
check that the correlation function in (\ref{g1}) involves only good
quark components, while in (\ref{gT}) necessarily one good and one bad
 components are involved. In the approach of \cite{KS70} only good
field components correspond to genuine partonic degrees of freedom,
while bad components are not dynamically independent and can
be eliminated through the equations of motion in favor of good
components and insertions of quark masses or gluon fields.
An important point is that this relation is nonlocal and involves quark
and gluon fields with different positions on the light-cone \cite{KS70}:
\ba{-+}
  q_-(\lambda n) = - i\int\! \frac{d\lambda'}{2\pi}\int\!\frac{ d\nu}{2\nu}\,
   e^{-i\nu(\lambda-\lambda')} \!\not\!n\!\not\!\!D_\perp(\lambda'n)\,
q_+(\lambda'n)\,,
\ea
where $D_\mu = \partial_\mu -igA_\mu$ is the covariant derivative.

Eq.~(\ref{-+}) states that degrees of freedom associated with the
bad component of the quark field in \re{gT} are, in fact, those of one quark and
one gluon. The structure function $g_T(x,Q^2)$ is, therefore, naturally
related to the quark-antiquark-gluon correlation function in the nucleon.
More precisely, $g_T(x)$ presents by itself only one special projection of this
more general three-particle distribution $D(\xi_1,\xi_2,\xi_3)$ that depends,
generally, on the momentum fractions $\xi_i$ carried by three partons.
It is this special projection, $g_T(x)$, that can be measured in deep inelastic
scattering with a transversely polarized target. On the other hand,
the scale dependence of the quark-antiquark-gluon distribution
function involves the ``full'' function $D(\xi_1,\xi_2,\xi_3)$ in a nontrivial
way \cite{BKL84,B-3g,DKM99} and the knowledge of one particular projection $g_T(x,Q_0^2)$
at a given value of $Q_0^2$ does not allow to predict $g_T(x,Q^2)$ at different
momentum transfers: a DGLAP-type evolution equation for $g_T(x,Q^2)$
in QCD does not exist or, at least, is not warranted. The reason is simply that
inclusive measurements in general do not provide complete information on the relevant
three-particle parton correlation function $D(\xi_1,\xi_2,\xi_3)$.

{}From the phenomenological point of view such situation is not satisfactory
since it would mean that one cannot relate results of the measurements of
$g_2(x)$ at different values of $Q^2$ to one another without model assumptions.
The theoretical challenge is, therefore, to find out whether the complicated
pattern of quark-gluon correlations can be reduced to a few effective degrees of
freedom. One has to look for meaningful approximations to the scale dependence
that introduce a minimum amount of nonperturbative parameters.

In particular, it was found by Ali, Braun and Hiller (ABH) \cite{ABH}
that the scale-dependence of the flavor-nonsinglet contribution
to $g_T(x,Q^2)$ simplifies dramatically in the limit of large number
of colors $N_c\to\infty$. To explain this result, it is convenient to
use the language of the Operator Product Expansion (OPE), see Sect.~2 for more
details. The statement of the OPE is that odd moments $n=3,5,\ldots$
of the structure function $g_2(x,Q^2)$ can be expanded in contributions of
multiplicatively renormalized local twist-3 quark-antiquark-gluon operators%
\footnote{Here we neglect the twist-2 contribution to $g_2(x)$.}
\ba{OPE}
 \int_0^1 \! dx\,x^{n-1} g_2(x,Q^2) &=&
   \sum_{k=0}^{n-3} C_{n-3}^k
   \left(\frac{\alpha_s(Q)}{\alpha_s(\mu)}\right)^{\gamma_{n-3}^k/b}
   \langle\!\langle O_N^k(\mu) \rangle\!\rangle ,
\ea
where  $C_{n-3}^k$ are the coefficients and
$\langle\!\langle O_{n-3}^k(\mu) \rangle\!\rangle$ reduced matrix elements
normalized at the scale $\mu$; $\gamma_{n-3}^k$ are the corresponding
anomalous dimensions that we assume are ordered with $k$:
$\gamma^0_{n-3} < \gamma^1_{n-3} < \ldots < \gamma^{n-3}_{n-3}$ for each $n$,
and $b=11N_c/3 -2n_f/3 $. The exact analytic expression for the lowest
anomalous dimension in the spectrum, $\gamma^0_{n-3}$, has been found in \cite{ABH}
and it was also noticed that the coefficient functions of all
other operators (with higher anomalous dimensions for each $n$) are suppressed
by powers of $1/N_c^2$. Thus, to the stated ${\cal O}(1/N_c^2)$ accuracy,
each moment of $g_2(x)$ involves a single nonperturbative parameter while
the complicated degrees of freedom related to quark-antiquark-gluon correlations
essentially decouple. The result can be reformulated  as a DGLAP-type evolution equation
\ba{AP:NS}
  Q^2 \frac{d}{d Q^2} \, g_2^{NS}(x,Q^2) &=& \frac{\alpha_s}{4\pi}
     \int_x^1\! \frac{dz}{z} P^{NS}(x/z) \,g_2^{NS}(z,Q^2)\,,
\nonumber\\
   P^{NS}(z) &=&
    \left[\frac{4C_F}{1-z}\right]_+
      +\delta(1-z)\left[C_F+\frac{1}{N_c}\left(2-\frac{\pi^2}{3}\right)\right]
      - 2C_F\,,
\end{eqnarray}
where $C_F=(N_c^2-1)/(2N_c)$ and we have included the $1/N_c^2$ corrections
calculated in \cite{BKM00}.

The present paper is devoted to the extension of this analysis to the
flavor-singlet sector in which case twist-3 three-gluon operators
have to be included. We calculate the leading one-loop ${\cal O}(\alpha_s)$
gluon contribution to the coefficient function and examine its
properties. We find that the one-loop coefficient function is such
that it mainly picks up the contribution of the twist-3 three-gluon
operator with the lowest anomalous dimension for each moment $N$.
The dominance of the lowest three-gluon ``state'' is observed both
for the logarithmic contribution $\sim \ln Q^2/\mu^2$ that reflects
mixing with the quark-antiquark-gluon operators, and the constant
term that gives rise to a ``genuine'' gluon contribution.

The result is very encouraging and allows to hope for the similar pattern of a
simplified evolution that mainly involves a single quark-antiquark gluon and a
single three-gluon parton distribution corresponding to the ``trajectories''
with the lowest anomalous dimension as important degrees of freedom, although
the reason for such a simplification is different. The properties of these
trajectories have been recently studied in \cite{B-3g}.

The outline of the paper is as follows. Sect.~2 is introductory and
reviews existing results on the OPE of the antisymmetric part of the
T-product of two electromagnetic currents to twist-3 accuracy.
The calculation of  the coefficient function  of twist-3 three-gluon operators.
is presented in Sect.~3 and its structure is elaborated upon in Sect.~4;
in Sect.~5 we summarize. Technical details on the twist separation
in gluon operators are presented in the Appendix.

\section{The Operator Product Expansion}
\setcounter{equation}{0}

As well known, the hadronic tensor which appears in the description
of deep inelastic  scattering of polarized leptons on polarized nucleons,
involves two structure functions
\ba{Wa}
  W^{(A)}_{\mu\nu} &=&\frac{1}{p\cdot q}
   \varepsilon_{\mu\nu\alpha\beta} q^\beta
\left\{ s^\beta g_1(x_B,Q^2) +\left[s^\beta- \frac{s\cdot q}{p\cdot q}p^\beta
\right]g_2(x_B,Q^2)\right\}\,,
\ea
where the nucleon spin vector is defined as
$s_\mu = \bar u(p,s)\gamma_\mu\gamma_5 u(p,s)$, $u(p,s)$ is the
nucleon spinor ($\bar u(p,s) u(p,s) =2 M$, $s^2=-4M^2$),
and is related to the imaginary part of the Fourier-transform of the
T-product of two electromagnetic currents, antisymmetrized over the
Lorentz indices:
\ba{T-product}
  W^{(A)}_{\mu\nu} &=& \frac{1}{\pi}\,{\rm Im}\, T^{(A)}_{\mu\nu}\,,
\nonumber\\
  i T^{(A)}_{\mu\nu} &=& \frac{i}{2} \!\int \!d^4x\,e^{iqx}
\langle p,s| T\{j_\mu(x/2)j_\nu(-x/2)-j_\nu(x/2)j_\mu(-x/2)\} |p,s\rangle\,.
\ea
We are going to examine the light-cone expansion of (\ref{T-product})
at $x^2\to 0$ and write down the answer in terms of nonlocal light-cone
operators of increasing twist, schematically
\ba{twistexpansion}
&&\frac{i}2 T\{j_\mu(x)j_\nu(-x)-j_\nu(x)j_\mu(-x) \} \stackrel{x^2\to 0}{=}
\nonumber
\\
&&
\qquad\qquad
\frac{i\varepsilon_{\mu\nu\alpha\beta}}{16\pi^2}
        \frac{\partial}{\partial x_\alpha}\left\{[C\ast O_\beta]^{\rm tw-2}
   + [C\ast O_\beta]^{\rm tw-3} +({\rm higher~twists})\right\},
\ea
where $C\ast O_\beta$ stands for the product (convolution) of the coefficient
functions and operators of the corresponding twist. This expression is explicitly
$U(1)$-gauge invariant, i.e. $\partial/\partial x_\mu T\{j_\mu(x)j_\nu(-x)\} =0$.

\subsection{Leading-order results}

\begin{figure}[t]
\centerline{\epsfxsize14.0cm\epsffile{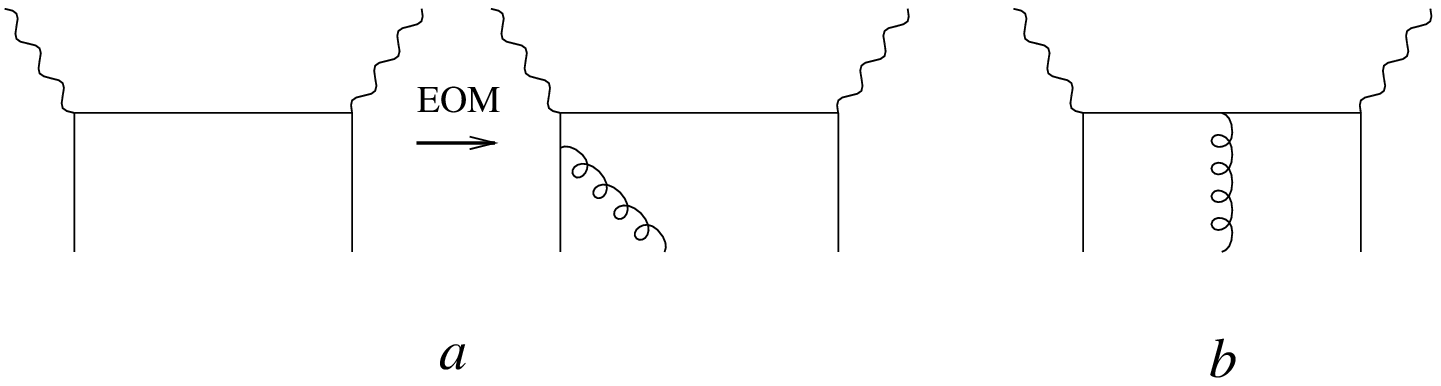}}
\caption[]{\small
Leading-order twist-3 contributions to the OPE of the T-product
of electromagnetic currents}
\label{OPEtree}
\end{figure}

To the leading order in the strong coupling, the OPE of the antisymmetric part of
the T-product in (\ref{twistexpansion}) can be written in a compact form as
\cite{BBunpublished}
\ba{OPE1}
 [C\ast O_\beta]^{\rm tw-2} &=& \frac{x_\beta}{x^4}
       \sum_{q=u,d,s,\ldots}\!\!\!e^2_q\,\,
\int_0^1\!\! du\,\, \bar q(ux)\!\not\!x\gamma_5 q(-ux)+(x\to -x)\,,
\nonumber\\
{} [C\ast O_\beta]^{\rm tw-3} &=& \frac{i}{2x^2}
        \sum_{q=u,d,s,\ldots}\!\!\!e^2_q\,\,
\int_0^1\!\! du\!\int^u_{-u} \!\!dv\,
\left[(u+v)\,S^+_\beta(u,v,-u)+(u-v)S^-_\beta(u,v,-u)\right]
\nonumber\\&&{}
+(x\to -x)\,,
\ea
where notation was introduced for light-cone nonlocal quark-gluon operators
\ba{Spm}
   S_\mu^\pm (a,b,c) &=& \frac12\bar q(ax)[ig\widetilde G_{\mu\nu}(bx)\pm
   gG_{\mu\nu}(bx)\gamma_5]x^\nu\!\not\!xq(cx)\,.
\ea
The dual gluon strength tensor is defined as $\widetilde G_{\mu\nu} =
\frac12\varepsilon_{\mu\nu\alpha\beta}G^{\alpha\beta}$ and we use  the
conventions $\gamma_5 = i\gamma_0\gamma_1\gamma_2\gamma_3$ and
$\varepsilon^{0123}=1$, see \cite{Okun}. To save space, here and below we do not
show the gauge factors connecting the quark (gluon) fields:
\be{[a,b]}
  [ax,bx] = {\rm P}\exp\left[ig\int_a^b\! du\,x_\mu A^\mu(ux)\right]\,.
\ee
The full twist-3 contribution \re{OPE1} to \re{twistexpansion}
is assembled from the Feynman diagrams shown in Fig.~\ref{OPEtree}.
The individual contributions of the two diagrams in Fig.~\ref{OPEtree} correspond
to the two possibilities to apply the derivative in \re{twistexpansion}:
$\partial/\partial x_\alpha
(S^\pm_\beta/x^2) = -2x_\alpha S^\pm_\beta/x^4 +1/x^2\partial
S^\pm_\beta/\partial x_\alpha$. The term  $\sim 1/x^4$ presents the contribution
of the the diagram in Fig.~\ref{OPEtree}a, rewritten in terms of quark-gluon
operators using equations of motion. The contribution $\sim 1/x^2$ corresponds to
the diagram with gluon emission from the hard propagator in
 Fig.~\ref{OPEtree}b and is necessitated by gauge invariance of (\ref{twistexpansion}).

Going over to matrix elements, we introduce the usual quark helicity
distributions $\Delta q(x_B) = q^\uparrow(x_B)-q^\downarrow(x_B)$
\be{deltaq}
\vev{p,s|\bar q(x)\not\!x\gamma_5 q(-x)|p,s}
=(sx)\int_{-1}^1\!d\xi\, \e^{2i\xi px}\,\Delta q(\xi,\mu^2)\,,
\ee
where $\mu$ refers to the normalization scale of the operator in
the l.h.s. and positive (negative) $\xi$ correspond to the
contribution of quarks (antiquarks)
$\Delta q(x_B) = q^\uparrow(x_B)-q^\downarrow(x_B)$,
$\Delta q(-x_B)
\equiv \Delta \bar q(x_B) = \bar q^\uparrow(x_B)-\bar q^\downarrow(x_B)$,
respectively.

Similarly, we define the twist-3  quark-antiquark-gluon
parton correlation functions as (cf. \cite{Jaffe83})%
\footnote{We always imply that $x^2$ can be put to zero in the operator
matrix elements. Under this condition, the nonlocal operators $S^\pm_\beta$
still contain a superposition of twist-3 and twist-4 terms \cite{GLR},
with the twist-4 terms being explicitly proportional to $x_\beta$.
The easiest way to separate the genuine twist-3 contribution
is to take the transverse projection $S^\pm_\beta \to S^\pm_\perp$,
 see also \cite{BB89,GLR} for explicit expressions.}
\be{Dq}
\vev{p,s| S_\mu^\pm(u,v,-u)|p,s}
= 2i(px)\left[s_\mu (px) - p_\mu (sx)\right] \!\int_{-1}^1\!\! {\cal
D}\xi\,
\e^{ipx[\xi_1 u + \xi_2 v - \xi_3 u]} D_q^\pm(\xi_1,\xi_2,\xi_3)
\ee
where
\be{Dxi}
 \int_{-1}^1\!\! {\cal D}\xi \equiv \int_{-1}^1\!\!d\xi_1d\xi_2d\xi_3
 \,\delta(\xi_1+\xi_2+\xi_3)\,.
\ee
The correlation functions $D_q^\pm(\xi_i)$ have the following
symmetry property:
\be{bose}
  D_q^\pm(\xi_1,\xi_2,\xi_3) = (D_q^\mp)^\ast(-\xi_3,-\xi_2,-\xi_1)
\ee
They are in general complex functions, but the imaginary parts do not
contribute to the structure functions and can be omitted \cite{Jaffe83}.

For further use, it is convenient to introduce a separate notation
for the nucleon matrix element of the specific combination of
quark-antiquark-gluon operators entering the OPE in (\ref{OPE1}):
\ba{Tq}
 \lefteqn{\hspace*{-1cm}\int^u_{-u} \!dv\,\vev{p,s|
\left[(u+v)\,S^+_\mu(u,v,-u)+(u-v)S^-_\mu(u,v,-u)\right]|p,s}=
\hspace*{5cm}{}}
\nonumber\\
&&\hspace*{3.5cm}{}=-i\left[s_\mu - p_\mu\frac{ (sx)}{(px)}\right]
\!\int_{-1}^1\!
 \!d\xi\, \e^{2i\xi upx}\,\, \Delta q_T(\xi,\mu^2)\,.
\ea
The function $\Delta q_T(x_B,Q^2)$ will play an important r\^ole in what follows.
It describes the momentum fraction distribution inside the nucleon of the
transverse spin and has the same support property as the parton distribution in
(\ref{deltaq}). However, in contrast with \re{deltaq}, it does not have
any probabilistic interpretation but can rather be expressed through the more
general three parton correlation functions $D_q^\pm(\xi_1,\xi_2,\xi_3)$, integrating
out the dependence on the gluon momentum fraction \cite{BKL84}.
Making a Fourier transformation of (\ref{OPE1}), taking imaginary part and
comparing with the definition of structure functions in (\ref{Wa}) one obtains
\be{g11}
    g_1(x_B,Q^2) = \frac{1}{2}\sum_q e^2_q\,
    [\Delta q(x_B,\mu^2 = Q^2)+\Delta q(-x_B,\mu^2=Q^2)]\,,
\ee
where $x_B= Q^2/(2pq)$ is the Bjorken variable, and \cite{BBunpublished,ABH}
\be{BCform}
    g_2(x_B,Q^2) = g_2^{WW}(x_B,Q^2) + \frac{1}{2}\sum_q e^2_q
      \int_{x_B}^1 \frac{dy}{y}
  [\Delta q_T(y, Q^2)+\Delta q_T(-y,Q^2)]
\ee
where
\be{gWW}
   g_2^{WW}(x_B,Q^2) = -g_1(x_B,Q^2) +
   \int_{x_B}^1 \frac{dy}{y}\, g_1(y,Q^2)
\ee
is the the familiar Wandzura-Wilczek contribution \cite{WW} and will
not be considered from now on.

Going over to the moments in \re{BCform} we obtain
\be{moments}
   \int_0^1\!dx_B \, x_B^{n-1} g_2(x_B,Q^2)
  = \frac{1}{2n} \sum_q e^2_q
   \int_0^1\!dx_B \, x_B^{n-1}
   [\Delta q_T(x_B,Q^2)+\Delta q_T(-x_B,Q^2)]\,.
\nonumber
\ee
For odd $n\ge 3$ the relevant integrals of the transverse spin
quark distributions $\Delta q_T$ are given by the reduced matrix elements
of local twist-3 operators that arise via the Tailor-expansion of the
nonlocal operators \re{Spm} at short distances $x_\mu \to 0$:
\be{localS}
{}[ S^\pm_\mu ]^k_N = \frac12\bar q (\derleft\cdot x)^k
    [ig\widetilde G_{\mu\nu}\pm  gG_{\mu\nu}\gamma_5]\not\! x\,x^\nu
\,(\derright\cdot x)^{N-k}q\,.
\ee
According to (\ref{Dq}), the reduced matrix elements
$\langle\!\langle\ldots \rangle\!\rangle$
of these operators
\be{vev}
\vev{p,s| [S_\mu^\pm]^k_N|p,s}
= 2(ipx)^{N+1}\left[s_\mu (px) - p_\mu (sx)\right]
\langle\!\langle [S^\pm]^k_N \rangle\!\rangle\,
\ee
are equal to moments of the quark-antiquark-gluon three-particle distribution
amplitude
\be{Dmoments}
  \langle\!\langle [S^\pm]^k_N \rangle\!\rangle =
\int_{-1}^1\!\! {\cal D}\xi\,
\xi_1^k \xi_3^{N-k} D_q^\pm(\xi_i)\,.
\ee
The symmetry relation (\ref{bose}) implies
\be{bose2}
  \langle\!\langle [S^\pm]^k_N \rangle\!\rangle^* = (-1)^N
 \langle\!\langle [S^\mp]^{N-k}_N \rangle\!\rangle\,.
\ee
Expanding Eq.~(\ref{Tq}) at short distances we obtain
\ba{OPE2}
   \int_{-1}^1\! d\xi \, \xi^{N+2} \Delta q_T(\xi) &=&
 2\sum_{k=0}^N (-1)^{N-k}
 \big[
      (k+1)\langle\!\langle [S^-]^k_N \rangle\!\rangle
   +(N-k+1)\langle\!\langle [S^+]^k_N \rangle\!\rangle
      \big]
\nonumber\\
 &=&  {4}\sum_{k=0}^N (-1)^{N-k}
      (k+1) \mbox{\rm Re}\, \langle\!\langle [S^-]^k_N \rangle\!\rangle
\nonumber\\
&=&  {4} \int_{-1}^1\!\! {\cal D}\xi\,
     \mbox{\rm Re}\, D_q^-(\xi_i)
   \frac{\partial}{\partial \xi_1} \frac{1}{\xi_1+\xi_3}
    \Big[ \xi_1^{N+2}-(-\xi_3)^{N+2}\Big]
      \,.
\ea
The last equality can also be obtained directly from the definition in
(\ref{Dq}), (\ref{Tq}). The following comments are in order.

We note the function $\Delta q_T(\xi)$ takes real values.
The expression in the last line of \re{OPE2} can be used for an analitic
continuation to $N\to-2$ and remains finite provided that the corresponding
integral of the $D-$function converges. This convergence, thus, presents a necessary
condition for the validity of the BC sum rule at $N=-2$.

At $N=-1$ an absence of a local twist-3 operator \re{localS} with dimension four
implies the constraint
\be{ETsumrule1}
 \int_{-1}^1\! d\xi \, \xi \,\Delta q_T(\xi) =
 \int_0^1\!dx_B \, x_B
   [\Delta q_T(x_B) - \Delta q_T(-x_B)] = 0\,.
\ee
This relation should be compared with the first moment of $g_2(x)$ given
by \re{moments} that involves the combination of the same distributions but
with a different C-parity
\be{ETsumrule2}
 \int_0^1\!dx_B \, x_B\, g_2(x_B,Q^2) =
     \frac14 \sum_q e^2_q
   \int_0^1\!dx_B \, x_B
   [\Delta q_T(x_B)+\Delta q_T(-x_B)]\,.
\ee
Vanishing of this integral (known as Efremov-Leader-Teryaev sum rule
\cite{ELT,AELreview}) is, therefore, not warranted by the OPE,
although its numerical value can be small since
the r.h.s. does not receive contribution from the valence
quarks.

\subsection{The scale dependence}

The dependence of the structure function $g_2(x,Q^2)$ on $Q^2$
is driven by the scale dependence of the distribution functions
$\Delta q_T(x,\mu^2)$. Going over to moments \re{OPE2}, it corresponds to the
renormalization-group scale dependence of the local operators
$[S^\pm]^k_N$. Similar to the familiar case of the helicity distributions
$\Delta q(x_B)$, one has to distinguish between the components with
different flavor symmetry as they have a different scaling behavior.
For instance, the flavor decomposition of the $u$-quark distribution
looks like
\ba{flavor}
 \Delta u_T(x_B) &=& \frac12\,\big(\Delta u_T-\Delta d_T\big)(x_B)
     +\frac16\,
            \big(\Delta u_T+\Delta d_T -2 \Delta s_T\big)(x_B)
\nonumber\\&+&
   \frac13\, \big(\Delta u_T+\Delta d_T + \Delta s_T\big)(x_B)\,.
\ea
Renormalization of flavor-nonsinglet contributions,
$\Delta q^{\rm NS}$, given by either $\Delta u - \Delta d$, or
$\Delta u + \Delta d- 2 \Delta s$
is simpler since they
do not mix with gluons. Still, the number of contributing operators
 rises linearly with $N$ \cite{SV82}. By an explicit calculation
one obtains, for the two lowest moments \cite{SV82,BKL84,ABH,Kodaira}
\ba{N=0,2}
\hspace*{-0.5cm}
 \frac14\int_{-1}^1\!d\xi \,\xi^2\,
    \Delta q_T^{\rm NS}(\xi, Q^2) &=&  L^{\gamma_0^0/b}
  \langle\!\langle S^0_0 (\mu^2)\rangle\!\rangle\,,
 \\
\hspace*{-0.5cm}
 \frac14\int_{-1}^1\!d\xi \,\xi^4\, \Delta q_T^{\rm NS}(\xi, Q^2) &=&
   L^{\gamma_2^0/b}
  \Big[
    0.807 \langle\!\langle S^0_2 (\mu^2)\rangle\!\rangle
  - 2.320 \langle\!\langle S^1_2 (\mu^2)\rangle\!\rangle
  + 2.894 \langle\!\langle S^2_2 (\mu^2)\rangle\!\rangle
 \Big]
\nonumber\\
&+&
 L^{\gamma_2^1/b}
  \Big[
    0.028 \langle\!\langle S^0_2 (\mu^2)\rangle\!\rangle
  - 0.014 \langle\!\langle S^1_2 (\mu^2)\rangle\!\rangle
  - 0.026 \langle\!\langle S^2_2 (\mu^2)\rangle\!\rangle
 \Big]
\nonumber\\
&+&
 L^{\gamma_2^2/b}
  \Big[
    0.165 \langle\!\langle S^0_2 (\mu^2)\rangle\!\rangle
  + 0.334 \langle\!\langle S^1_2 (\mu^2)\rangle\!\rangle
  + 0.132 \langle\!\langle S^2_2 (\mu^2)\rangle\!\rangle
 \Big]
\nonumber
\ea
where we used
$\langle\!\langle S^k_N (\mu^2)\rangle\!\rangle \equiv
\mbox{\rm Re}\,\langle\!\langle [S^-]^k_N (\mu^2)\rangle\!\rangle$
as a shorthand and $L = \alpha_s(Q^2)/\alpha_s(\mu^2)$.
Anomalous dimensions are equal to \cite{ABH}
\be{lowdim}
 \gamma_0^0 = 8.5\,,\quad
 \gamma_2^0 = 10.89\,,\quad
 \gamma_2^1 = 13.71\,,\quad
 \gamma_2^2 = 16.15\,.
\ee
Eq.~(\ref{N=0,2}) illustrates the main difficulty: since the number of
contributing operators prolifiterates with $N$, so does the number of
independent nonperturbative parameters $\langle\!\langle S^k_N
(\mu^2)\rangle\!\rangle$. On the other hand, Eq.~(\ref{N=0,2}) reveals a
remarkable pattern: coefficient functions in front of the operators with higher
anomalous dimensions are much smaller than those with the lowest anomalous
dimension. This structure is not accidental, but related to a dramatic
simplification of the renormalization-group evolution of flavor-nonsinglet
operators in the large$-N_c$ limit. As was found in \cite{ABH}, the small
coefficients in (\ref{N=0,2}) are in fact suppressed by powers of $1/N_c^2$ and
one obtains in the limit $N_c\to\infty$
\ba{largeNc1}
\int_{-1}^1\!d\xi \,\xi^2\, \Delta q_T^{\rm NS}(\xi, Q^2) &\stackrel{N_c\to\infty}{=}&
   L^{\gamma_2^{\rm NS}/b}
  \Big[
     \langle\!\langle S^0_2 (\mu^2)\rangle\!\rangle
  - 2 \langle\!\langle S^1_2 (\mu^2)\rangle\!\rangle
  + 3 \langle\!\langle S^2_2 (\mu^2)\rangle\!\rangle
 \Big]
\nonumber\\
&=& L^{\gamma_2^{\rm NS}/b}\int_{-1}^1\!d\xi \,\xi^2\,
\Delta q_T^{\rm NS}(\xi,\mu^2)\,,
\ea
so that the scale evolution of the second moment of $\Delta q_T^{\rm NS}$ involves the same moment.
The similar phenomenon takes place for arbitrary $N$:
\be{largeNc2}
\int_{-1}^1\!d\xi \,\xi^N\,
\Delta q_T^{\rm NS}(\xi, Q^2) \stackrel{N_c\to\infty}{=}
  L^{\gamma_N^{\rm NS}/b} \int_{-1}^1\!d\xi \,\xi^N\,
\Delta q_T^{\rm NS}(\xi, \mu^2)\,.
\ee
Here $\gamma_N^{\rm NS}$ is the lowest anomalous dimension in the spectrum
of flavor-nonsinglet twist-3 operators. It is known analytically
in the large$-N_c$ limit \cite{ABH} and the $1/N_c^2$ corrections
have been recently calculated using the large$-N$ expansion in \cite{BKM00}:
\ba{gammaNS}
 \gamma_N^{\rm NS} &=& N_c \left(2 \psi(N+3) +2 \gamma_E
       +\frac{1}{N+3}-\frac12\right)
\nonumber\\
&-&\frac{2}{N_c}\left(\ln(N+3)+\gamma_E+\frac34 -\frac{\pi^2}{6}
    + {\cal O}\left(1/N^2\right)\right)
 +{\cal O}\left(\frac{1}{N_c^4}\right).
\ea
Here  $\psi(x)=d\ln\Gamma(x)/d x$ stands for the Euler
$\psi$-function.
Eq.~(\ref{largeNc2}) together with (\ref{gammaNS}) are equivalent to
the DGLAP evolution equation in (\ref{AP:NS}).

The scale dependence of the flavor-singlet distribution
$\Delta q_T^{\rm S}=\Delta u_T+\Delta d_T + \Delta s_T$ differs from
the above in three aspects. First, the mixing matrices of the relevant
quark-antiquark-gluon operators receive extra terms related to the possibility of
quark-antiquark annihilation. Second, they mix in addition with an entirely new
and equally big set of three-gluon operators. Third, the three-gluon operators
themselves contribute to the OPE of the T-product of the electromagnetic currents
starting order $\alpha_s$. Concerns have been raised (see e.g. \cite{Ji99}) that
in particular the last contribution does not have a simple structure and will
spoil any ABH--type approximation in the singlet case. We begin, therefore, with
the corresponding calculation.

\section{The gluon contribution to the structure functions}
\setcounter{equation}{0}

\begin{figure}[t]
\centerline{\epsfxsize8.0cm\epsffile{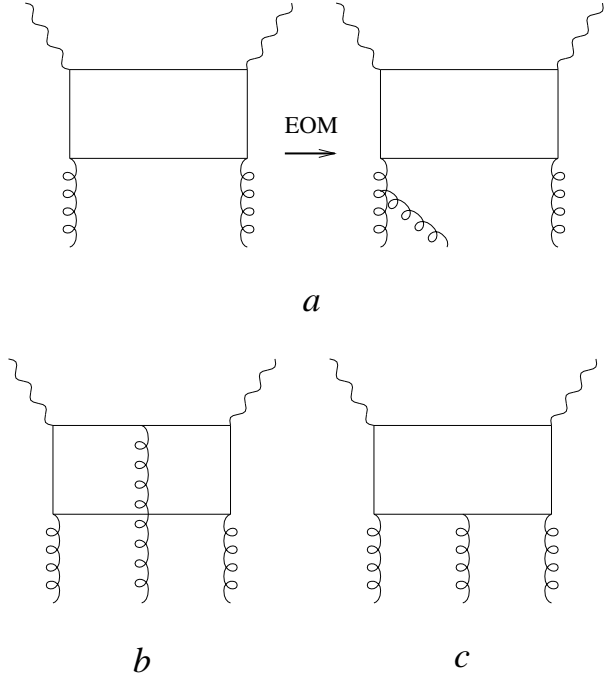}}
\caption[]{\small
 Leading-order twist-3 gluon contribution to the OPE of
the T-product of electromagnetic currents.}
\label{OPEgluon}
\end{figure}

The leading-order gluon contribution to the T-product of two electromagnetic
currents in (\ref{T-product}) is described by the box diagram shown in
Fig.~\ref{OPEgluon}. Its calculation can be easily done using the background
field approach of Ref.~\cite{BB89}. Namely, considering outgoing gluons as
classical background Yang-Mills fields we calculate the box diagram replacing the
free quark propagators by propagators in an external fields. Then, the
antisymmetric part of the T-product in (\ref{T-product}) is given by
\ba{polar}
    \Pi^{(A)}_{\mu\nu}(x,-x) &=&
\frac12 T\{j_\mu(x)j_\nu(-x)\} - (\mu\leftrightarrow\nu)
\nonumber\\
&=& \frac12 \sum_{q=u,d,s,\ldots}\!\!\!e^2_q\,
\Tr\Big[\gamma_\mu S(x,-x)\gamma_\nu S(-x,x)\Big]
- (\mu\leftrightarrow\nu)\,,
\ea
where $\Tr$ denotes the trace over both color and spinor indices, and $S(x,-x)$
is the quark propagator in a background gluon field
\be{back}
  S(x,-x) = \langle x| \frac{1}{-i\!\not\!\partial -g\!\not\!\!A}|-x\rangle\,.
\ee
For $x^2\to 0$ the propagator $S(x,-x)$ exhibits light-cone singularities that
one handles using the dimensional regularization with $d=4-2\epsilon$ and
$\epsilon <0$. Then, expanding $S(x,-x)$ in powers of the deviation from the
light-cone and retaining contributions of gluon operators up to twist-3 we find
\be{S(x,-x)}
  (4\pi)^{d/2} S(x,-x) =
-\frac{\Gamma(d/2)}{(-x^2)^{d/2}}\, S_0
            -\frac{\Gamma(d/2-1)}{(-x^2)^{d/2-1}}\, S_1
            - \frac{\Gamma(d/2-2)}{(-x^2)^{d/2-2}}\,S_2+\ldots
\ee
First two terms in the light-cone expansion of the propagator are known
\ba{prop}
 S_0(x,-x) &=& \not\!x[x,-x]\,,
\nonumber\\
 S_1(x,-x) &=&
 \frac12\int_{-1}^1\!\!du\,
  \Bigg\{ [x,ux] \Big[g\widetilde G_{x\alpha}(ux)\gamma_\alpha \gamma_5
     -iu gG_{x\alpha}(ux)\gamma_\alpha\Big][ux,-x]
\nonumber\\
 &+& \!\not\!x \!\int_{-1}^u\!\!dv\,(1-u)(1+v) [x,ux]gG_{x\alpha}(ux)
[ux,vx] gG_{x\alpha}(vx)[vx,-x]\Bigg\}
\ea
and the further ones  can be calculated using the technique described in
Appendix~A of \cite{BB89}. Here and below we use a
shorthand notation $G_{x\alpha}=
G_{\mu\alpha}^a x_\mu t^a$ with $t^a$ being the generators of fundamental (quark)
representation of the $SU(N_c)$.

Omitting  the disconnected (gluon field independent) contribution $\sim
1/(-x^2)^d$, one obtains
\be{sin-anal}
\Pi^{(A)}_{\mu\nu}(x,-x) =\!\!
\sum_{q=u,d,s,\ldots}\!\!\!\!e^2_q\,
\frac{\Gamma^2(d/2-1)}{(4\pi)^d (-x^2)^{d-2}}
\left[A_{\mu\nu}(x,-x)+\frac{d-2}{d-4}B_{\mu\nu}(x,-x)\right]
+ {\cal O}\!\left(\frac1{(-x^2)^{d-3}}\right),
\ee
where
\ba{AB}
A_{\mu\nu}&=&\frac12\Tr\Big[\gamma_\mu S_1(x,-x)\gamma_\nu S_1(-x,x)\Big]-
(\mu\leftrightarrow\nu)\,,
\\
B_{\mu\nu}&=&\frac12\Big\{
\Tr\Big[\gamma_\mu S_2(x,-x)\gamma_\nu S_0(-x,x)\Big] +
\Tr\Big[\gamma_\mu S_0(x,-x)\gamma_\nu S_2(-x,x)\Big]\Big\}-
(\mu\leftrightarrow\nu)\,.
\nonumber
\ea

Several comments are in order. First, the expression in (\ref{sin-anal}) defines
the most singular, $\sim 1/x^4$ as $d\to 4$, contribution to the light-cone
expansion of the T-product (\ref{twistexpansion}). This contribution  alone
suffices to determine the coefficient functions of the twist-2 and twist-3 gluon
operators since less singular contributions of the same twists can be uniquely
restored from matching \re{sin-anal} into the general $U(1)$-gauge invariant
expression (\ref{twistexpansion}), cf. the discussion in Sect.~2.1. Second,
notice that the first term in (\ref{sin-anal}) is analytic and the second  is
singular in the limit $d\to4$. These two terms correspond to the two distinct
integration regions in the quark momentum in the loop (see Fig.~2) $k^2 \sim
1/(-x^2)\sim Q^2 $ and  $k^2\ll Q^2$, respectively.
The first term, coming from large momenta, determines the
one-loop ${\cal O}(\alpha_s)$ coefficient function of gluon operators at a hard
scale of order $Q$, while the second term will be interpreted as a tree-level
quark coefficient function times the one-loop evolution (mixing) into gluons.
Finally, one can convince oneself that the traces in (\ref{sin-anal}) can be
calculated in dimension $d=4$. This is obvious for the first term, and can be
shown for the second, the reason being that calculation of diagrams of the type
shown in Fig.~2 does not involve contraction of Lorentz indices of
$\gamma$-matrices.

Calculating (\ref{sin-anal}) we shall assume the translation invariance of
$\Pi^{(A)}(x,-x)$ along the light-cone\footnote{That is,
neglect contributions of the operators containing total derivatives
that have vanishing forward matrix elements.}.
In addition, we shall impose the equations of motion for gluon
fields, $[D^\mu, G_{\mu\nu}(x)]=0$, which amounts to putting external gluons in
the box diagram (see Fig.~2) on their mass-shell. Then, using the explicit
expression for $S_1(x,-x)$ given in (\ref{prop}) one obtains
\ba{anal1}
A_{\mu\nu}(x,-x) &=&
 8 g^2 x_\mu\! \!\int_0^1\!\! du\,u\bar u\, \Tr\Big\{
   G_{x\xi}(ux)G_{\nu\xi}(-ux) - G_{x\xi}(-ux)G_{\nu\xi}(ux)\Big\}
\nonumber\\&+&
   4 i g^2 x_\mu\! \!\int_0^1\!\! du\,\bar u^2
     \!\int_{-u}^{u}\!\!dv\,
\Big[
  \left(1\!+\!v\!-\!\frac{\bar u}{3}\right) O_{\nu}(u,-u,v)
+\left(1\!-\!v\!-\!\frac{\bar u}{3}\right) O_{\nu}(v,u,-u)
\nonumber\\&&{}
\hspace*{3.5cm}+\frac{2}{3}\bar u O_{\nu}(u,v,-u)\Big]
-(\mu\leftrightarrow\nu)\,,
\ea
where we used $\bar u \equiv 1-u$ etc.  and introduced a notation
\be{GGGop}
  O_{\nu}(u,v,t) =
 g\Tr\Big\{\Big[G_{x\alpha}(ux),G_{x\nu}(vx)\Big]G_{x\alpha}(tx)\Big\}
 = \frac{ig}{2} f^{abc}G^a_{x\alpha}(ux) G^b_{x\nu}(vx)G^c_{x\alpha}(tx)\,.
\ee
In the last relation the gauge factors between the gluon fields in the adjoint
representation are implied.

The calculation of the second, singular contribution in (\ref{sin-anal})
is more tedious. After considerable algebra we obtain, however,
 an equally simple expression
\ba{singular}
B_{\mu\nu}(x,-x)&=& 16 g^2 \! \!\int_0^1\!\! du\,u\bar u\, \Tr\Big\{
   G_{x\mu}(ux)G_{x\nu}(-ux\Big\}
\nonumber\\
   &-& 4 i g^2 x_\mu\! \!\int_0^1\!\! \!du\,\bar u^2
     \!\!\int_{-u}^{u}\!\!dv
\Big[\!
  \left(3\!+\!v\!-\!\frac53{\bar u}\right) O_{\nu}(u,-u,v)
+\left(3\!-\!v\!-\!\frac53{\bar u}\right) O_{\nu}(v,u,-u)
\nonumber\\&&{}\hspace*{3.5cm}
     +\frac{2}{3}(2+u)\, O_{\nu}(u,v,-u)\Big]
-(\mu\leftrightarrow\nu)\,.
\ea
Substituting Eqs.~\re{GGGop} and \re{singular} into \re{sin-anal} we finally
obtain the leading order expression for the antisymmetric part of the T-product
of electromagnetic currents that takes into account both twist-2 and twist-3
contributions. However, in order to match \re{AB} into the general OPE form
\re{twistexpansion} we have to separate the different twists.

\subsection{Twist separation}
The two-gluon operators in (\ref{anal1}) and (\ref{singular})
contain, generically, contributions of both twist-2 and twist-3.
The separation of twists corresponds in this case to the
separation of the terms of different symmetry and can be done using the trick
described in \cite{BB89}. The twist-2 part of the two-gluon operator in
(\ref{anal1}) can be written as (see Appendix A for details)
\be{trick1}
  \Big[\Tr\left\{G_{x\xi}(x)G_{\nu\xi}(-x)\right\}\Big]^{\rm tw-2} =
  \int_0^1\!du\,u\frac{\partial}{\partial x_\nu}
 \Tr\left\{G_{x\xi}(ux)G_{x\xi}(-ux)\right\}.
\ee
Similarly, neglecting the irrelevant operators proportional to the equations of
motion and total derivatives, one obtains the twist-3 contribution as
\ba{trick3}
\Big[\Tr\left\{G_{x\xi}(x)G_{\nu\xi}(-x)\right\}\Big]^{\rm tw-3}\!\!\!{} &=&
 i\!\!\int_0^1\!\!\!du\!\int_{-u}^u\!\!\!dv\,
 \Tr\Bigg\{ (1+u^2)\, [G_{x\xi}(ux), G_{x\xi}(vx)] G_{x\nu}(-ux)
\nonumber\\&&\hspace*{1.5cm}
        -(1-uv)\,  [G_{x\xi}(ux), G_{x\nu}(vx)] G_{x\xi}(-ux)
\nonumber\\&&\hspace*{1.5cm}
        + (1-u^2)\, [G_{x\nu}(ux), G_{x\xi}(vx)] G_{x\xi}(-ux)\Bigg\}.
\ea
Applying (\ref{trick1}) we find that the twist-2 two-gluon contribution to
(\ref{anal1}) in fact cancels out, and the remaining twist-3 part can be
rewritten using (\ref{trick3}) as
\ba{trick4}
\lefteqn{\hspace*{-2cm}[\Tr\left\{G_{x\xi}(ux)G_{\nu\xi}(-ux)-
 G_{x\xi}(-ux)G_{\nu\xi}(ux)\right\}]^{\rm tw-3}=}
\nonumber\\
 &=&  -2i\!\int_0^u\!ds\!\int_{-s}^s\!dt\,
\Big\{O_{\nu}(s,-s,t) + O_{\nu}(s,t,-s) + O_{\nu}(t,s,-s)
\Big\}.
\ea
Substitution of this relation into \re{anal1} yields
\ba{anal2}
[A_{\mu\nu}]^{\rm tw-2}&=& 0\,,
\nonumber\\{}
[A_{\mu\nu}]^{\rm tw-3} &=& -4ig^2 x_\mu\!\!\int_0^1\!\! du\,\bar u^2\!
\!\int_{-u}^{u}\!\!\!dv\,\Big\{
 (u-v)\,O_{\nu}(u,-u,v) + 2u \,O_{\nu}(u,v,-u)\mbox{\hspace*{2.4cm}}
\nonumber\\&&{}\hspace*{3.6cm}
+(u+v)\,O_{\nu}(v,u,-u)\Big\}
- (\mu\leftrightarrow\nu)\,.
\ea
The separation of twists in (\ref{singular}) is equally simple. To this end we
note that (in dimension $d=4$)
\be{dualform}
 G_{x\mu}(ux)G_{x\nu}(-ux)-G_{x\nu}(ux)G_{x\mu}(-ux)=
\varepsilon_{\mu\nu\rho\sigma}x^\sigma G_{x\alpha}(ux)
\,\widetilde G_{\rho\alpha}(-ux)
\ee
and the relations (\ref{trick1}) -- (\ref{trick3}) remain true to the claimed
accuracy if one of the gluon strength-tensors is substituted by its dual
counterpart. We obtain
\ba{sin}
[B_{\mu\nu}]^{\rm tw-2}&=&-16g^2\varepsilon_{\mu\nu\rho\sigma}x^\sigma
\frac{\partial}{\partial x_\rho}
\!\int_0^1\!\!du\,(u\ln u +u\bar u)
\Tr\{ G_{x\xi}(ux)\widetilde G_{x\xi}(-ux)\}
\,,
\nonumber\\{}
[B_{\mu\nu}]^{\rm tw-3} &=&
 16i g^2 x_\mu\! \!\int_0^1\!\! du\! \!\int_{-u}^{u}\!\!dv\,\Big\{
 (\bar u u+\mbox{$\frac14$} \bar u^2 +u\ln u)
\Big[v\, O_{\nu}(v,u,-u)+ u\, O_{\nu}(u,v,-u)
\nonumber\\&-&
 v\, O_{\nu}(u,-u,v)\Big]
 -\frac{1}{12}\bar u^2(u+2)\Big[O_{\nu}(u,-u,v)+O_{\nu}(u,v,-u)+
O_{\nu}(v,u,-u)\Big]\Big\}
\nonumber\\&&{}
  - (\mu\leftrightarrow\nu)\,.
\ea

\subsection{Twist-2: Results}

Substituting (\ref{sin}), (\ref{anal2}) into (\ref{sin-anal}),
subtracting the (collinear) singularities in the
$\overline{\rm MS}$ scheme%
\footnote{Note that the (non-renormalized) tree-level coefficient function in
coordinate space contains a notrivial $d-$dependence corresponding
to the quark propagator $\Gamma(d/2)\!\!\not{\!\!\!x}/(-4\pi x^2)^{d/2}$.
Making the $\overline{\rm MS}-$subtraction in coordinate space one has to
keep this factor in dimension $d$ in the counter-term. Alternatively,
one can make the Fourier transform first, and then subtract the
divergencies in the usual way.
One can check that the renormalized coefficient functions
obtained in both ways are indeed related to each other by a 4-dimensional
Fourier transformation.}
and combining with the leading-order result
in (\ref{OPE1}) , we obtain the
twist-2 contribution
\ba{renorm-2}
 [C\ast O_\beta]^{\rm tw-2} &=& \frac{x_\beta}{x^4}
       \sum_{q=u,d,s,\ldots}\!\!\!e^2_q\,\,
\int_0^1\!\! du\,\, \Bigg\{
\Big[\bar q(ux)\!\not\!x\gamma_5 q(-ux)+(x\to -x)\Big]_{\mu^2_{\overline{\rm MS}}}
\\
&+&\frac{4\alpha_s}{\pi}
\Big( \ln(-x^2\mu^2_{_{\overline{\rm MS}}})+2\gamma_E\Big)
(u\ln u +u\bar u) \Tr\{ G_{x\xi}(ux)\widetilde G_{x\xi}(-ux)\}\Bigg\}\,,
\nonumber
\ea
where the subscript $[\ldots]_{\mu^2}$ indicates the normalization scale of the
operator. Note simplicity of the answer: the entire gluon contribution
can be eliminated by choosing the proper scale of the quark operator
$\mu^2_{\overline{\rm MS}} = 1/(-x^2e^{2\gamma_E})$. This property
is lost in the momentum space since after the Fourier transformation
contributions of different light-cone separations get mixed.

Going over to the matrix elements, we introduce the usual
gluon  helicity distribution~\cite{Man91,BB91}
\be{Deltag}
\vev{p,s|\Tr\Big\{G_{x\alpha}(x)\widetilde G_{x\alpha}(-x)\Big\}|p,s}
=\frac{i}{4}(sx)(px)
\int_{-1}^1\!d\xi\, \e^{2i\xi px}\,\xi \Delta g(\xi,\mu^2)\,.
\ee
Note that $\Delta g(\xi)=\Delta g(-\xi)$.

Moments of the structure functions (\ref{Wa}) are obtained by the expansion of
the T-product (\ref{T-product}) in momentum space in powers of $\omega =
-2(pq)/q^2$, $Q^2=-q^2$  in the unphysical region $\omega\to 0$, and matching to
the corresponding expansion in terms of structure functions:
\be{dis-rel}
T_{\mu\nu}^{(A)}= - 4\varepsilon_{\mu\nu\alpha\beta}\frac{q_\alpha}{q^2}
\sum_{n=1,3,\ldots} \omega^{n-1}
\left\{ s^\beta g_1(n,Q^2) +\left[s^\beta- \frac{s\cdot q}{p\cdot q}p^\beta
\right]g_2(n,Q^2)\right\},
\ee
where the moments are defined as (for any function $f$)
\be{moms}
f(n,Q^2) = \int_0^1 \!dx_B\, x_B^{n-1}\, f(x_B,Q^2)
\,.
\ee

Taking the Fourier transform of \re{twistexpansion} by using
\re{renorm-2} and matching the obtained expression into \re{dis-rel}
we obtain after some algebra
\ba{mom-g1}
 &&g_1(n,Q^2) = \frac12
\sum_{q=u,d,s,\ldots}\!\!\!e^2_q\,
\Bigg\{\Delta q(n,\mu^2_{\overline{\rm MS}}) +
\Delta \bar q(n,\mu^2_{\overline{\rm MS}})
\nonumber\\
&&
\qquad \qquad\qquad+
\frac{\alpha_s}{2\pi}
\frac{n-1}{n(n+1)}\Delta g(n,\mu^2)\left[
\ln\frac{Q^2}{{\mu^2_{\overline{\rm MS}}}}-\psi(n)-1-\gamma_E\right]
\Bigg\},
\nonumber\\
&&[g_2(n,Q^2)]^{\rm tw-2}=-\frac{n-1}n\,g_1(n,Q^2)\,,
\ea
in accord with the Wandzura-Wilczek relation \re{gWW}. Comparing the first
expression in (\ref{mom-g1}) with the general expression
\be{general1}
g_1(n,Q^2) = \frac12\!\!
\sum_{q=u,d,s,\ldots}\!\!\!\!e^2_q\,\Bigg\{
\Delta q(n,\mu^2)+ \Delta \bar q(n,\mu^2) +
\Big[\gamma_{qg}(n)\ln\frac{Q^2}{\mu^2}+C_g(n)\Big]
\Delta g(n;\mu^2)+\ldots\Bigg\}
\ee
we find the anomalous dimension and the gluon coefficient function as
\ba{gammaQG}
  \gamma_{qg}(n) &=& \frac{\alpha_s}{2\pi}\frac{n-1}{n(n+1)}\,,
\nonumber
\\
  C_g(n) &=&
  -\frac{\alpha_s}{2\pi}\frac{n-1}{n(n+1)}\Big[\psi(n)+1+\gamma_E\Big]\,.
\ea
Going over from the moments to the momentum fraction representation,\\
$\gamma_{qg}(n)=\int_0^1 dx\, x^{n-1} P_{qg}(x)$ and $C_g(n)=\int_0^1 dx\,
x^{n-1} C_g(x)$, we get
\ba{PQG}
  P_{qg}(x) &=& \frac{\alpha_s}{2\pi}(2x-1)\,,
\nonumber
\\
   C_g(x) &=&
  -\frac{\alpha_s}{2\pi}\Big[(2x-1)\ln\frac{x}{1-x}+4x-3\Big]\,.
\ea
Expressions in \re{gammaQG} and \re{PQG} are in agreement with the well-known
results, see e.g.  \cite{T2}.

\subsection{Twist-3: Results}

The twist-3 contribution to the T-product \re{sin-anal} comes from \re{anal2} and
\re{sin}. To cast it into the $U(1)-$gauge invariant form \re{twistexpansion}
we notice that
\ba{dual1}
 x_\mu O_\nu(u,v,t)-x_\nu O_\mu(u,v,t) = \varepsilon_{\mu\nu\alpha\beta}x^\alpha
\widetilde O_\beta(u,v,t) + {\cal O}(x^2)\,,
\ea
where the operators $O_\nu$ were defined in \re{GGGop} and
\be{Odual}
  \widetilde O_{\beta}(u,v,t)
 = \frac{ig}{2} f^{abc}G^a_{x\alpha}(ux)
   \widetilde G^b_{x\beta}(vx)G^c_{x\alpha}(tx)\,.
\ee
Then, combining together Eqs.~\re{anal2} and \re{sin},
subtracting the collinear
singularities in the $\overline{\rm MS}$ scheme (see previous footnote) and
comparing with the general structure of the OPE in (\ref{twistexpansion}) we
obtain:
\ba{OPE-tw3}
&& \hspace*{-5mm}[C\ast O_\beta]^{\rm tw-3}=
\nonumber\\
&& \frac{i}{2x^2}
        \sum_{q=u,d,s,\ldots}\!\!\!e^2_q\,\,
\int_0^1\!\! du\!\int^u_{-u} \!\!dv\,\Bigg\{
(u+v)\,S_\beta(u,v,-u)
+ (u-v)  S_\beta(-u,v, u)
\nonumber\\
&& + \bar u^2 \frac{\alpha_s}{\pi}
  \Big[(u+v)\widetilde O_{\beta}(v,u,-u)+2u \widetilde O_{\beta}(u,v,-u)
+ (u-v)\widetilde O_{\beta}(u,-u,v)\Big]
\\
&& +\frac{4\alpha_s}{\pi}
\left(\ln(-x^2\mu^2_{\overline{\rm MS}})+2\gamma_E+1\right)\Big[
 (\bar u u+\mbox{$\frac14$} \bar u^2 +u\ln u)
\left[v\, \widetilde O_{\beta}(v,u,-u)+ u\,\widetilde O_{\beta}(u,v,-u)\right.
\nonumber\\&&
\left.
- v\, \widetilde O_{\beta}(u,-u,v)\right]
 -\frac{1}{12}\bar u^2(u+2)\left[\widetilde O_{\beta}(u,-u,v)+\widetilde O_{\beta}(u,v,-u)+
\widetilde O_{\beta}(v,u,-u)\right]\Big]\Big\}_{\mu^2_{\overline{\rm MS}}},
\nonumber
\ea
where the subscript ${\mu^2_{\overline{\rm MS}}}$ indicates the normalization
point of nonlocal quark and gluon operators. Here we introduced the C-even
quark-gluon operator
\be{qGq-even}
 S_\mu(u,v,-u) = S_\mu^{+}(u,v,-u)+S_\mu^{-}(-u,v,u)
\ee
so that
\be{C-parity}
 S_\mu(-u,v,u) =  S_\mu^{-}(u,v,-u)+S_\mu^{+}(-u,v,u)\,.
\ee
The following comments are in order.

The twist-3 gluon contribution in
\re{OPE-tw3} has two parts. The lengthy expression in the last two lines can in
fact be eliminated by the choice of scale in the quark operator in the first
line: $\mu^2_{\overline{\rm MS}}= 1/(-x^2
\e^{2\gamma_E+1})$. As a nontrivial check of our calculation, we have verified
that our answer \re{OPE-tw3} is in agreement with the renormalization group
equation for the twist-3 operator $S_\mu(u,v,-u)$~\cite{BB89,Muller}
\ba{QGrenorm}
\lefteqn{
 [ S_\beta(u,v,-u)]_{\mu_2^2} \,\,{=}\,\,
[S_\beta(u,v,-u)]_{\mu_1^2}
\,\,-\,\frac{\alpha_s}{2\pi}
\ln\frac{\mu_2^2}{\mu_1^2}
\! \int_{-u}^u\!\!ds\!\int_{-u}^s\!\! dt\, (2u)^{-3}}
\nonumber\\&\times&
 \!\!\Big\{[2u(s-t)+4(u-s)(t+s)]\widetilde O_{\beta}(s,v,t)
- 2|u|(s-t)[\widetilde O_{\beta}(s,t,v)-\widetilde O_{\beta}(v,s,t)]\Big\}.
\mbox{\hspace*{0.8cm}}
\ea
In addition, the expression in the second line in \re{OPE-tw3} defines a
`genuine' twist-3 gluon coefficient function that cannot be eliminated by the
scale choice in the quark operator. This expression is surprisingly simple and
can be cast in the form similar to that of the tree-level contribution of the
quark operators, Eq.~\re{OPE1}:
\be{OPE3}
[C\ast O_\beta]^{\rm tw-3}_{\rm gluon} = \frac{i}{2x^2}
        \sum_{q=u,d,s,\ldots}\!\!\!e^2_q\,\,
\int_0^1\!\! du\!\int^u_{-u} \!\!dv\,
\left[(u+v)\,{\cal G}_\beta(u,v,-u)
- (u-v){\cal G}_\beta(-u,v,u)\right],
\ee
where
\be{cal-S}
{\cal G}_\beta(u,v,-u) = -\frac{\alpha_s}{\pi}
\bar u^2 \left[\widetilde O_{\beta}(v, u, - u) + \widetilde O_{\beta}(u,v,-u)
\right].
\ee

The nucleon matrix element of \re{OPE-tw3} defines the twist-3 gluon
correlation function similar to \re{Dq}\footnote{cf. the footnote to \re{Dq}}
\be{Dg}
\vev{p,s| \widetilde O_{\mu}(u,v,-u)|p,s}
= -2(px)^2\left[s_\mu (px) - p_\mu (sx)\right] \!\int_{-1}^1\!\! {\cal D}\xi\,
\e^{ipx[\xi_1 u + \xi_2 v - \xi_3 u]} D_g(\xi_1,\xi_2,\xi_3)
\ee
with the integration measure given by \re{Dxi}.
Since $\widetilde O_{\mu}(u,v,-u)=-\widetilde O_{\mu}(-u,v,u)$ according to
the definition \re{Odual}, the correlation function $D_g(\xi_i)$
is antisymmetric to the interchange of the first and the third argument:
\be{bose-g}
D_g(\xi_1,\xi_2,\xi_3)=- D_g(\xi_3,\xi_2,\xi_1)\,.
\ee
Expanding the nonlocal operator \re{Odual} over local twist-3 gluon operators
\be{localG}
[G_\mu]_N^k = \frac{i}2 g f^{abc} G_{x\alpha}^a\,(\derleft\cdot x)^k\,
\widetilde G_{x\mu}^b\,(\derright\cdot x)^{N-1-k}\,G_{x\alpha}^c
\ee
and defining the reduced matrix elements $\langle\!\langle\ldots
\rangle\!\rangle$ of these operators
\be{vev1}
\vev{p,s| [G_\mu]^k_N|p,s}
= 2(ipx)^{N+1}\left[s_\mu (px) - p_\mu (sx)\right]
\langle\!\langle [G]^k_N \rangle\!\rangle\,
\ee
we obtain the moments of the gluon three-particle distribution amplitude
(\ref{Dg}) as
\be{Dg-moments}
  \langle\!\langle [G]^k_N \rangle\!\rangle =
\int_{-1}^1\!\! {\cal D}\xi\,
\xi_1^k \xi_3^{N-1-k} D_g(\xi_i)
\ee
with  $0\le k \le [N/2]-1$. The symmetry  \re{bose-g} implies that
$\langle\!\langle [G]^k_N\rangle\!\rangle=-\langle\!\langle [G]^{N-1-k}_N
\rangle\!\rangle$ and, therefore,
 the number of independent gluon matrix elements is equal to $[N/2]$.

Finally, substituting \re{OPE-tw3} into \re{twistexpansion} and taking the
Fourier transform \re{T-product}, we match the result into the expansion
\re{dis-rel} to obtain the moments of the structure function for $n=1,3,\ldots$
\ba{otvet}
 [g_2(n,Q^2)]^{\rm tw-3} &=& \frac12  \sum_{q=u,d,s,\ldots}\!\!\!e^2_q\,\,
\frac{4}{n}\int_{-1}^1\!{\cal D}\xi\,
\Bigg\{ D_q(\xi_i,{\mu_{\overline{\rm MS}}^2})\, \Phi_n^{q}(\xi_1,\xi_3) \,
\\[3mm]
&&{}\hspace*{-0.5cm}
+ \frac{\alpha_s}{4\pi} \frac{D_g(\xi_i,{\mu_{\overline{\rm MS}}^2})}{n+1}
\Bigg[ \Phi_n^{g}(\xi_i)+\Omega_n^{qg}(\xi_i)
\Bigg(\ln\frac{Q^2}{\mu_{\overline{\rm MS}}^2}
-\psi\left(n\right)-\!\gamma_E\!-1\!\Bigg) \Bigg]\Bigg\}\,,
\nonumber\ea
where, as usual, applicability to the lowest moments relies on the 
extra assumption about the high-energy asymptotics of the cross section, 
alias the assumption that the corresponding dispersion relation does not 
involve extra subtraction constants.  
Here $D_q(\xi_i)$ is the distribution function corresponding to the $C-$even
quark-gluon operator \re{qGq-even}
\be{D=2ReD}
D_q(\xi_i,{\mu_{\overline{\rm MS}}^2}) = {\mbox{\rm
Re}}\,D^-_q(\xi_i;{\mu_{\overline{\rm MS}}^2})\,,
\ee
the quark coefficient function is defined as 
\ba{psi-n}
\Phi_n^{q}(\xi_1,\xi_3)&=&\frac{\partial}{\partial \xi_1}
            \frac{\xi_1^{n-1}-(-\xi_3)^{n-1}}{\xi_1+\xi_3}
\ea
and it has already appeared in \re{OPE2}. The gluon coefficient
functions can be expressed in terms of $\Phi_n^{q}$ as
\ba{CD}
&&\Phi_n^g(\xi_i) =
\left[\Phi_{n-1}^{q}(\xi_1,\xi_3)+\Phi_{n-1}^{q}(\xi_1,-\xi_1-\xi_3)\right]
+(\xi_1\leftrightarrow -\xi_3)\,,
\\[3mm]
&&\Omega_n^{qg}(\xi_i) =
\left(1+\frac{2}{n(n-2)}\right) \Phi_n^g(\xi_i)
 +\frac{2(n-1)}{n(n-2)}\left[\Phi_{n-1}^{q}(-\xi_1-\xi_3,\xi_3)
 +\Phi_{n-1}^q(\xi_1+\xi_3,-\xi_1)\right]\!.
\nonumber
\ea
Explicit expressions for a few first moments read:
\ba{low-n}
&&
\Phi_{3}^g=\Omega_{3}^{qg} = 0\,,
\nonumber\\
&&
\Phi_{5}^g=5\,(\xi_{{1}}-\xi_{{3}})\,,\quad
\Omega_{5}^{qg} = {\frac {31}{5}}\,(\xi_{{1}}-\xi_{{3}})\,,
\\
&&
\Phi_{7}^g=14\,\left (\xi_{{1}}^3-\xi_{{3}}^3\right )\,,
\quad
\Omega_{7}^{qg}= \frac27\,\left (59\,\xi_{{1}}^3+6\xi_1^2 \xi_3-6\xi_1 \xi_3^2
-59\,\xi_{{3}}^3\right )\,.
\ea

The expressions \re{OPE-tw3}, (\ref{OPE3}) and
\re{otvet} for the one-loop gluon contribution to the antisymmetric part of the
T-product of two electromagnetic currents (\ref{twistexpansion}) and the
structure function $g_2(x_B,Q^2)$ present the main result of this section.

\section{Properties of the twist-3 contribution}
\setcounter{equation}{0}

According to \re{otvet},  the moments of the structure function are given by
integrals of the quark-gluon and three-gluon distribution functions, $D_q(\xi_i)$ and
$D_g(\xi_i)$, respectively, over momentum fractions of partons with the weights defined
by the coefficient functions $\Phi_n^q(\xi_i)$, $\Phi_n^g(\xi_i)$ and $\Omega_n^{qg}(\xi_i)$.

The quark coefficient function \re{psi-n} vanishes at $n=1$ and for higher
moments $\Phi_n^q$ is a homogenous polynomial in momentum fractions $\xi_i$
of degree $n-3$. For the gluon coefficient functions, Eqs.~\re{CD}, one
finds that $\Omega^{qg}_{n=1}(\xi_i) = \Phi^g_{n=1}(\xi_i) = 0$ and, therefore,
the gluon contribution to the first moment of $g_2(x,Q^2)$ vanishes, in agreement
with the BC sum rule, provided that the three-gluon distribution function $D_g(\xi_i)$
does not have additional singularities. Moreover, in contrast with the quark
coefficient function,
$\Omega^{qg}_{n=3}(\xi_i)=\Phi^g_{n=3}(\xi_i) = 0$, and gluon contribute to the
structure function starting from $n=5$th moment. For $n\ge 5$ the coefficient
functions $\Omega^{qg}_{n}$ and $\Phi^g_{n}$ are homogenous polynomials in
$\xi_i$ of degree $n-4$.

The explicit expressions for the lowest moments \re{otvet} look as follows
\ba{explicit}
&&\int_0^1 dx\, x^2 g_2^{\rm tw-3}(x,Q^2) = \frac23\sum_q e_q^2 \int_{-1}^1 {\cal D} \xi\,
D_q(\xi_i;Q^2)\,,
\\
&&
\int_0^1 dx\, x^4 g_2^{\rm tw-3}(x,Q^2) = \frac25\sum_q e_q^2 \int_{-1}^1 {\cal D} \xi\,
\\
&& \hspace*{20mm}\times\left[ (3 \xi_1^2 - 2 \xi_1 \xi_3 + \xi_3^2)D_q(\xi_i;Q^2)
-\frac{\alpha_s}{\pi} \frac{847}{1440}(\xi_1-\xi_3)D_g(\xi_i;Q^2)\right]\,,
\nonumber\\
&&
\int_0^1 dx\, x^6 g_2^{\rm tw-3}(x,Q^2) = \frac27\sum_q e_q^2 \int_{-1}^1 {\cal D} \xi\,
\\
&& \hspace*{20mm}\times\left[ (5 \xi_1^4 - 4 \xi_1^3 \xi_3 + 3\xi_1^2\xi_3^2
-2\xi_1\xi_3^3+ \xi_3^4)D_q(\xi_i;Q^2)
\right.\nonumber\\&& \hspace*{32mm}\left.
-\frac{\alpha_s}{\pi}\left(\frac{3091}{2240}(\xi_1^3-\xi_3^3)
+\frac{207}{1120}(\xi_1^2\xi_3-\xi_1\xi_3^2)\right)D_g(\xi_i;Q^2)\right]\,.
\nonumber
\ea
Using \re{Dmoments} and \re{Dg-moments} we express the moments in
terms of reduced quark and gluon matrix elements
\ba{mom-ope}
&&\int_0^1 dx\, x^2 g_2^{\rm tw-3}(x,Q^2) = \frac23\sum_q e_q^2
\langle\!\langle S^0_0 (Q^2)\rangle\!\rangle,
\\
&&
\int_0^1 dx\, x^4 g_2^{\rm tw-3}(x,Q^2)
\\
&&\qqqquad= \frac25\sum_q e_q^2\left[
    \langle\!\langle S^0_2 (Q^2)\rangle\!\rangle
- 2 \langle\!\langle S^1_2 (Q^2)\rangle\!\rangle
+ 3 \langle\!\langle S^2_2 (Q^2)\rangle\!\rangle
+ \frac{\alpha_s}{\pi} \frac{847}{720}\langle\!\langle G^0_2 (Q^2)\rangle\!\rangle\right],
\nonumber
\\
&&
\int_0^1 dx\, x^6 g_2^{\rm tw-3}(x,Q^2)
\\
&&\qqqquad= \frac27\sum_q e_q^2\left[
    \langle\!\langle S^0_4 (Q^2)\rangle\!\rangle
- 2 \langle\!\langle S^1_4 (Q^2)\rangle\!\rangle
+ 3 \langle\!\langle S^2_4 (Q^2)\rangle\!\rangle
- 4 \langle\!\langle S^3_4 (Q^2)\rangle\!\rangle
+ 5 \langle\!\langle S^4_4 (Q^2)\rangle\!\rangle
\right.\nonumber \\
&&\qqqquad\qqqquad
\left.
+\frac{\alpha_s}{\pi}\left(\frac{3091}{1120}\langle\!\langle
G^0_4 (Q^2)\rangle\!\rangle+\frac{207}{560}\langle\!\langle
G^1_4 (Q^2)\rangle\!\rangle\right)\right].
\nonumber
\ea
We would like to stress that the relations \re{otvet}, \re{explicit} and
\re{mom-ope} take into account the leading order contribution of the
three-gluon operators and they do not include ${\cal O}(\alpha_s)-$corrections
to the coefficient functions of quark-antiquark-gluon operators.
The latter corrections have been recently calculated in \cite{Ji}
using a different operator basis.

We can make one further step  and define the momentum fraction distribution of
the transverse spin carried by gluons in the nucleon by the expression similar to
\re{Tq}:
\ba{Tg}
 \lefteqn{\hspace*{-2cm}\int^u_{-u} \!dv\,\vev{p,s|
\Big[(u+v)\widetilde O_{\mu}(v,u,-u)+2u\, \widetilde O_{\mu}(u,v,-u)
+ (u-v)\widetilde O_{\beta}(u,-u,v)\Big]
|p,s}=
\hspace*{5cm}{}}
\nonumber\\
&&\hspace*{3.5cm}{}=2\left[s_\mu (px) - p_\mu (sx)\right]
\!\int_{-1}^1\!
 \!d\xi\, \e^{2i\xi upx}\,\,\xi\, \Delta g_T(\xi,\mu^2)\,,
\ea
so that (cf. \re{OPE2})
\ba{OPEglue}
   \int_{-1}^1\! d\xi \, \xi^{n-1} \Delta g_T(\xi) &=&
  2 \int_{-1}^1\!\! {\cal D}\xi\,
     \,\Phi_n^g(\xi_i)\, D_g(\xi_i)
\ea
and $\Delta g_T(\xi)=\Delta g_T(-\xi)$.
It is easy to see  that the first contribution in the square brackets
in \re{otvet} can be easily rewritten in terms of
$\Delta g_T(n,Q^2) = \int_0^1\!
d\xi\, \xi^{n-1} \Delta g_T(\xi)$. Unfortunately, the two coefficient functions
$\Phi_n^g(\xi_i)$ and $\Omega_n^{qg}(\xi_i)$ do not coincide and, therefore, the
full gluon contribution to \re{otvet} cannot
be, strictly speaking, reduced to the contribution of $\Delta
g_T(\xi)$ alone, as a yet another manifestation of the fact that we are dealing
with a three-particle problem.

Our main observation is that such a reduction can, nevertheless, provide
a reasonable approximation to the moments $g_2(n,Q^2)$
\be{approx}
\int_{-1}^1\!\! {\cal D}\xi\,D_g(\xi_i;Q^2)\,
\Omega_n^{qg}(\xi_i)\approx c(n)\,\int_{-1}^1\!\!
{\cal D}\xi\,D_g(\xi_i;Q^2)\,\Phi_n^g(\xi_i)\,,
\ee
with  the coefficient of proportionality $c(n)$ that is {\em independent\/}
on the momentum transfer $Q^2$.
In this case,
moments of the structure function \re{otvet} can be expressed in terms
of the quark and gluon distributions as
\ba{otvet-2}
&&\int_0^1\!dx_B \, x_B^{n-1} g_2(x_B,Q^2)\approx \frac{1}{2n} \sum_q e^2_q
   \int_0^1\!dx_B \, x_B^{n-1}
\\
&&\qqqquad\qquad
\times
   \left[\Delta q_T(x_B,Q^2)+\Delta q_T(-x_B,Q^2)
   +C_g^{\rm tw-3}(n)\,\Delta g_T(x_B,Q^2)\right].
\nonumber
\ea
where $\Delta q_T$ and $\Delta g_T$ were defined in \re{Tq} and \re{OPEglue},
respectively, and
\be{c-tw3}
C_g^{\rm tw-3}(n)=-\frac{\alpha_s}{2\pi}\frac1{n+1}[(\psi(n)+\gamma_E+1)\,c(n)-1]\,.
\ee
The reason for \re{approx} to hold is that, as we shall argue below,
the both projections of the three-gluon distribution defined in
the l.h.s and the r.h.s of
Eq.~(\ref{approx}) can be identified to a good numerical accuracy with
the contribution
of the three-gluon multiplicatively renormalizable operator with the lowest anomalous
dimension. To be more precise, this statement refers to the ``purely gluonic''
operator, defined without taking into account the mixing with the
quark-antiquark-gluon sector. We have checked that the mixing between
three-gluon and quark-antiquark-gluon operators does not
modify the result significantly; a detailed analysis will be
presented elsewhere~\cite{2}.

To justify this, we note that in a ``purely gluonic'' sector an
arbitrary multiplicatively renormalizable twist-3 three-gluon operator
${\cal O}_{N,\alpha}$
can be characterized  by the coefficients in its expansion over the
basis of operators $[G_\mu]^k_N$ defined in (\ref{localG})
\be{4-co}
   {\cal O}_{N,\alpha} = \sum_{k=0}^{[N/2]-1} w^k_{N,\alpha}\, [G_\mu]^k_N
   = \frac12 \sum_{k=0}^{[N/2]-1} w^k_{N,\alpha}\,
   \left([G_\mu]^k_N-[G_\mu]^{N-1-k}_N\right)
\ee
or, equivalently, by a characteristic polynomial
\be{4-cf}
 W_{N,\alpha}(\xi_1,\xi_3)=\frac12\sum_{k=0}^{[N/2]-1}w^k_{N,\alpha}
 \left(\xi_1^k \xi_3^{N-1-k}- \xi_1^{N-1-k}\xi_3^k\right)\,.
\ee
The subscript $0\le \alpha \le [N/2]-1$ enumerates the operators and we assume,
for definiteness, that the operators ${\cal O}_{N,\alpha}$ are ordered in such
a way that a smaller $\alpha$ corresponds to a lower anomalous dimension.
With this definition, it follows from (\ref{Dg-moments})
 that the reduced matrix element of a multiplicatively renormalizable operator
${\cal O}_{N,\alpha}$ is given by a weighted integral of the
three-particle gluon distribution:
\be{4-w}
  \langle\!\langle {\cal O}_{N,\alpha} \rangle\!\rangle =
\int_{-1}^1\!\! {\cal D}\xi\,
  \,W_{N,\alpha}(\xi_i)\, D_g(\xi_i)\,.
\ee
For the purpose of our discussion, we assume that the expansion coefficients $w^k_{N,\alpha}$
in \re{4-co} are calculated by an explicit diagonalization of the mixing matrix given in
\cite{BKL84} so that the polynomials $W_{N,\alpha}(\xi_i)$ are known functions.

\begin{figure}[th]
\centerline{\epsfxsize7.5cm\epsffile{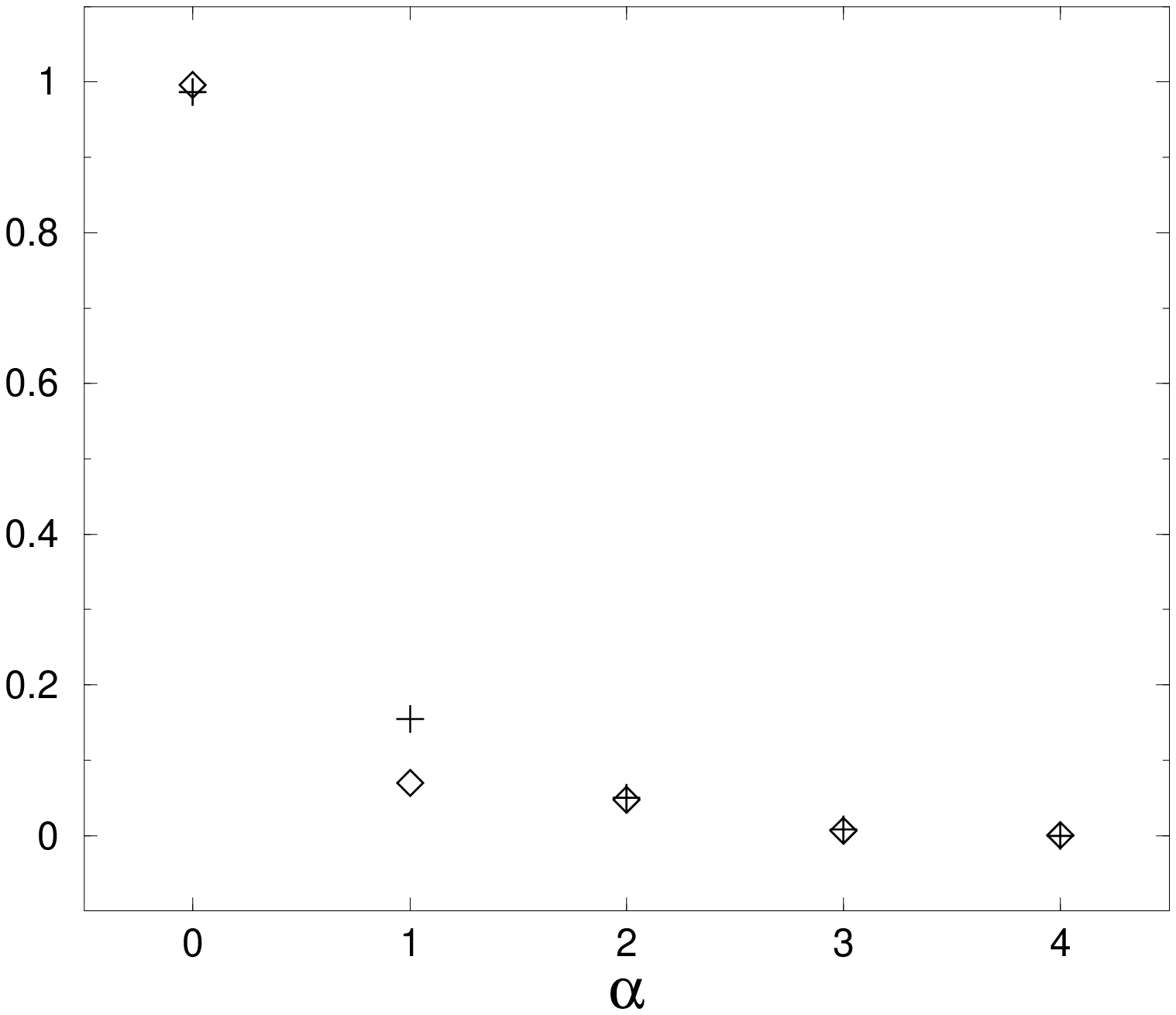}\quad
\epsfxsize7.5cm\epsffile{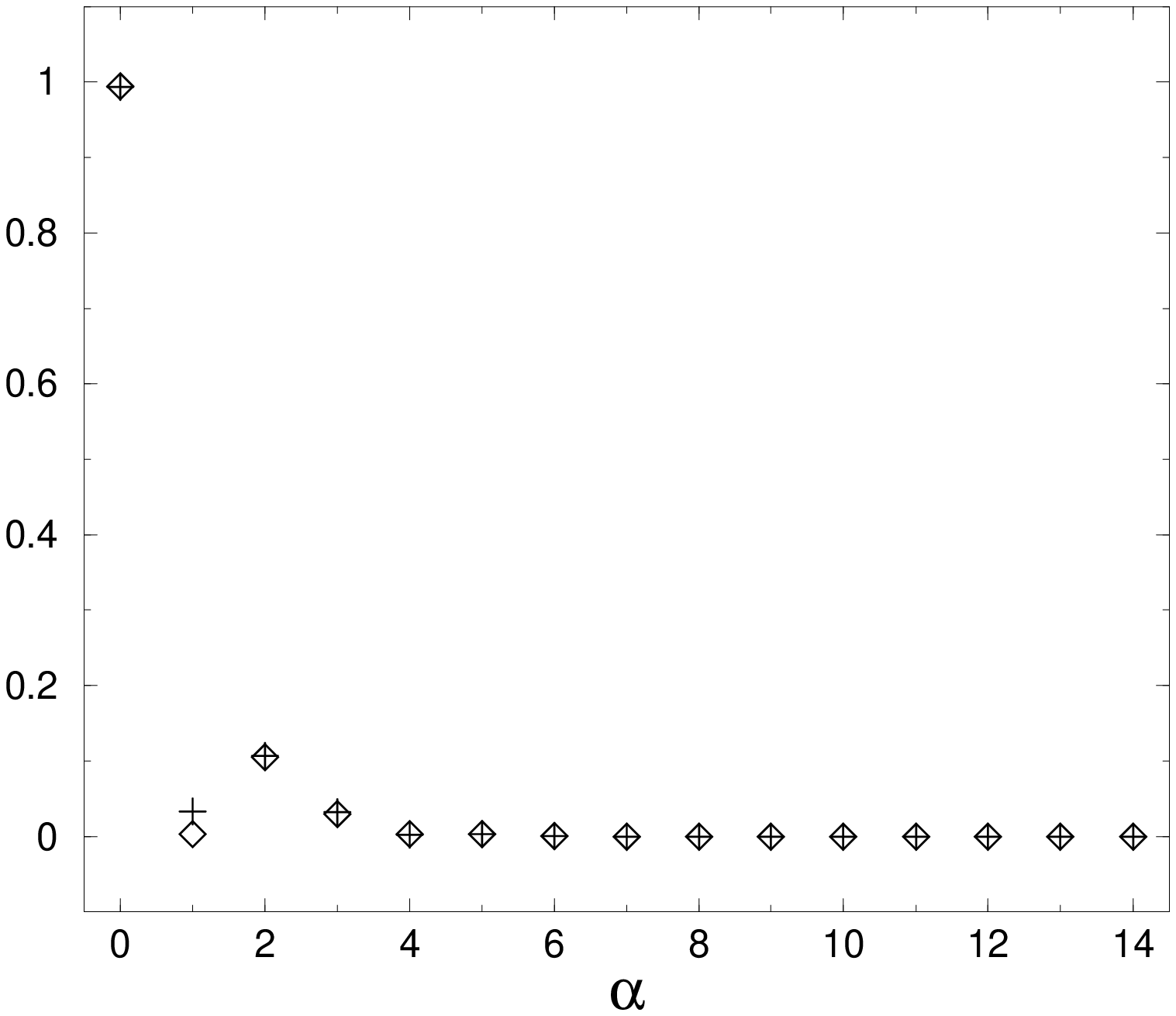}}
\caption[]{\small Conformal projection (see text) of the coefficient functions
$\Phi^g_n(\xi_i)$ (crosses) and
$\Omega^{qg}_n(\xi_i)$ (diamonds) on the contributions of multiplicately
renormalizable three-gluon
operators enumerated by an integer variable $\alpha$ in the order of the
increasing anomalous dimension. Mixing with quark-antiquark-gluon
operators is neglected. The two plots correspond
to the moments $n=13$ (right) and $n=33$ (left), respectively.}
\label{low-level-dominance}
\end{figure}
To prove our assertion, we have to
show that the coefficient functions $\Phi_n^g(\xi_i)$ and
$\Omega^{qg}_n(\xi_i)$
are numerically close to $W_{N,\alpha=0}(\xi_i)$,
at least for sufficiently large values of $n=N+3$, or,
equivalently, the both sides of \re{approx} receive a dominant
contribution from $\langle\!\langle {\cal O}_{N,\alpha=0} \rangle\!\rangle$.
This is not straightforward since the characteristic polynomials
$W_{N,\alpha}(\xi_i)$ for different $\alpha$ are not mutually
orthogonal with respect to any simple weight function, the reason being
that the mixing matrices \cite{BKL84} are not symmetric.

In order to make a meaningful comparison we use the conformal
symmetry that allows to rewrite the mixing matrices in a different
basis such that they become hermitian (see \cite{BKM00,BDKM} for details).
In the present context, the idea is that the conformal symmetry
allows for a unique analytic continuation of the functions
$W_{N,\alpha}(\xi_i)$, $\Phi_N^g(\xi_i)$ and $\Omega^{qg}_n(\xi_i)$ defined on the plane
$\xi_1+\xi_2+\xi_3=0$ to the full three-dimensional space of
the momentum fraction variables $\xi_i$
\be{4-su2}
      W_{N,\alpha}(\xi_1,\xi_3) \ \Rightarrow\ \widehat{W}_{N,\alpha}(\xi_1,\xi_2,\xi_3)
\ee
etc., such that $W_{N,\alpha}(\xi_1,\xi_3)=\widehat{W}_{N,\alpha}(\xi_i)$ for
$\xi_1+\xi_2+\xi_3=0$ and the functions $\widehat W_{N,\alpha}(\xi_i)$
corresponding to different multiplicatively renormalizable operators
are mutually orthogonal with respect to the conformal scalar product
\be{int-rep}
\vev{\widehat W_{N,\alpha}|\widehat W_{N,\beta}}=
\int_0^1 [d\xi]\,\xi_1^2 \xi_2^2 \xi_3^{2}\,
\widehat W_{N,\alpha}(\xi_i)\widehat W_{N,\beta}(\xi_i)
\sim \delta_{\alpha\beta}\,,
\ee
where the integration measure is defined as
\be{int-measure}
[d\xi]=d\xi_1 d\xi_2 d\xi_3\,\delta(\xi_1+\xi_2+\xi_3-1)\,.
\ee
Using this method, we can expand $\widehat\Phi_n^g(\xi_i)$ and
$\widehat\Omega_n^{qg}(\xi_i)$ over the set of orthogonal polynomials
$\widehat W_{N,\alpha}(\xi_i)$ to arrive after reduction to
$\xi_1+\xi_2+\xi_3=0$ at
\be{4-expand}
  \Phi_n^g(\xi_i) = \sum_{\alpha=0}^{[N/2]-1}
\frac{\vev{\widehat \Phi_{n}^g|\widehat W_{N,\alpha}}}
    {\|\widehat W_{N,\alpha}\|^2} W_{N,\alpha}(\xi_i)
=\|\widehat \Phi_{n}^g\| \sum_{\alpha=0}^{[N/2]-1}\phi_{n,\alpha}
\frac{W_{N,\alpha}(\xi_i)}
    {\|\widehat W_{N,\alpha}\|}
\ee
with $n=N+3$ and similarly for $\Omega_n^{qg}(\xi_i)$. Here, the norm is defined in
a natural way as $\|\Psi\|^2 = \vev{\Psi|\Psi}$.

The results of our calculations of the (normalized) expansion coefficients
\be{4-norm}
\phi_{n,\alpha} =\frac{\vev{\widehat \Phi_{n}^g|\widehat W_{N,\alpha}}}
    {\|\widehat \Phi_{n}^g\|\cdot\|\widehat W_{N,\alpha}\|}\,,
\qquad
\omega_{n,\alpha} =
    \frac{\vev{\widehat \Omega_{n}^{qg}|\widehat W_{N,\alpha}}}
    {\|\widehat \Omega_{n}^{qg}\|\cdot\|\widehat W_{N,\alpha}\|}
\ee
corresponding to the functions $\Phi_{n}^g$ and $\Omega_{n}^{qg}$,
respectively,
are shown in Fig.~\ref{low-level-dominance} for $n=13$ and $n=33$.
It is clearly seen that
the sum \re{4-expand} is dominated by the single
contribution $\alpha=0$ of the three-gluon operator with the lowest
anomalous dimension. The quality of
such approximation is improving for larger moments $n$.
Moreover, we find that, in
accord with \re{approx}, the two functions $\Phi_n^g(\xi_i)$ and
$\Omega_n^{qg}(\xi_i)$ are, in fact, close to each other with
the coefficients $c(n)$ in \re{approx} and \re{c-tw3} given by
\be{c-n}
c(n) =\frac{\vev{\widehat \Omega_{n}^{qg}|\widehat W_{N,0}}}
{\vev{\widehat \Phi_{n}^g|\widehat W_{N,0}}}\,.
\ee
Since the contribution of $\Omega_n^{qg}(\xi_i)$ to \re{otvet}
reflects the mixing of three-gluon with quark-antiquark gluon
operators, this is an indication that the observed dominance of
the lowest three-gluon state is not obstructed by this mixing.

\begin{figure}[t]
\centerline{\epsfxsize8.0cm\epsffile{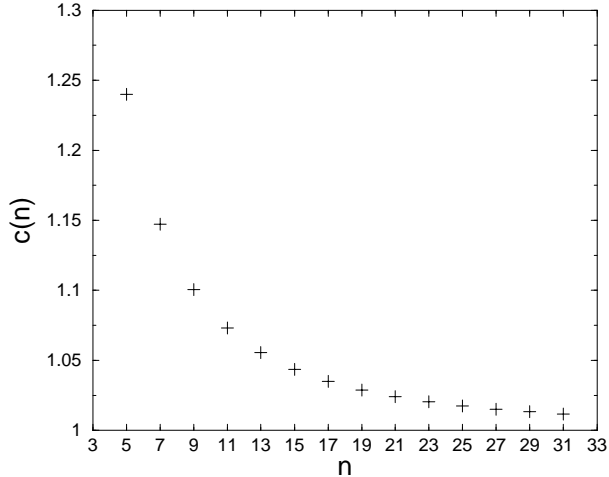}}
\caption[]{\small
Conformal projection $c(n)$
of the full gluon contribution, Eq.~\re{c-n}, onto the transverse spin gluon distribution
defined in \re{Tg}.}
\label{cn}
\end{figure}

{}From Eq.~\re{CD} it is easy to see that $c(n)\to 1$ for large moments $n$
whereas from \re{low-n} it follows that $c(5) = 31/25$. The values
of $c(n)$ for other (odd) moments $n<25$ are shown in Fig.~\ref{cn}.
The $n$-dependence is very smooth (for odd $n$) and can be approximated as
\be{c-param}
 c(n)= 1+11.11\,\frac{1}{n^2}- 25.74\,\frac{1}{n^3}\,.
\ee
Analytic expressions for $c(n)$ can be worked out in the large-$N_c$
limit and will be presented in \cite{2}.

\section{Conclusions}
We have presented a detailed calculation of the three-gluon
twist-3 contribution to the flavor-singlet structure function $g_2(x,Q^2)$
to the one-loop accuracy. The result is encouraging as it indicates that
the gluon coefficient function is close to that of the three-gluon
operator with the lowest anomalous dimension, at least for large moments.
This allows to hope for a simplified description of the scale dependence
of $g_2(x,Q^2)$ in terms of DGLAP equations, similar as for the
structure functions of leading twist.
The construction of this approximation requires a more detailed analysis
of the evolution equations for the flavor-singlet twist-3 operators
and will be given in a forthcoming publication~\cite{2}.

\subsection*{Acknowledgements}
This work was supported in part by the DFG, project Nr. 920585 (A.M.)
and by the EU networks ``Training and Mobility of Researchers'',
contracts EBR FMRX--CT96--008 (V.B.) and  FMRX--CT98--0194 (G.K.).

\subsection*{Note added}

When this work was in preparation, the work \cite{Ji2} appeared
with the calculation of the singlet twist-3 coefficient
functions using a different approach. The relation of this calculation
to our result is not obvious because of different operator basis.
It appears that the answer for the $n=5$ moment of $g_2(x)$ given in
\re{explicit} is in agreement with the appropriate projection
of the coefficient function calculated in \cite{Ji2}. We thank
A.~Belitsky for the correspondence on this topic.

\appendix
\renewcommand{\theequation}{\Alph{section}.\arabic{equation}}
\setcounter{table}{0}
\renewcommand{\thetable}{\Alph{table}}

\section{Appendix}

\setcounter{equation}{0}

The short-distance expansion of the nonlocal two-gluon operator looks like
\footnote{Throughout this Appendix we assume that gauge factors in
the adjoint representation are inserted in between the gluon fields.}
\be{2-gluon}
G_{x\alpha}(x)G_{\nu\alpha}(-x)=
\sum_{n=1}^\infty \frac1{(n-1)!} x_{\mu_1}
 x_{\mu_3}\ldots  x_{\mu_n}
G_{\mu_1\alpha} \deriv_{\mu_2} \ldots \deriv_{\mu_n} G_{\nu\alpha}(0)
\nonumber
\ee
The twist-2 contribution to the r.h.s.\ originates from the local operators
completely symmetric and traceless with respect to their Lorentz indices
\ba{twist-2-gluon}
\lefteqn{\hspace*{-1cm}
[G_{x\alpha}(x)G_{\nu\alpha}(-x)]^{\rm tw-2}=}
\nonumber\\
&&\sum_{n=1}^\infty
\frac1{(n-1)!(n+1)}
\frac{\partial}{\partial x_\nu}\left(x_{\mu_1}x_{\mu_2}\ldots  x_{\mu_n}
x_{\mu_{n+1}}\right) G_{\mu_1\alpha} \deriv_{\mu_2} \ldots \deriv_{\mu_n}
G_{\mu_{n+1}\alpha}(0)\,,
\ea
where $\deriv_{\mu}= \derleft_\mu-\derright_\mu$. It is straightforward to verity
that the same expression can be obtained through the following integral
representation
\be{twist-sep}
[G_{x\alpha}(x)G_{\nu\alpha}(-x)]^{\rm tw-2} = \int_0^1du\,u \frac{\partial}{\partial
x_\nu}G_{x\alpha}(ux)G_{x\alpha}(-ux)\,.
\ee
Subtracting the twist-2 contribution from the nonlocal two-gluon operator we
obtain the relation
\ba{A1}
\lefteqn{\hspace*{-2cm}
G_{x\alpha}(x)G_{\nu\alpha}(-x) - \int_0^1\!udu\,\frac{\partial}{\partial x_\nu}
   G_{x\alpha}(ux)G_{x\alpha}(-ux) =}
\nonumber\\
&=& \int_0^1\!udu\,x_\rho \left[
  \frac{\partial}{\partial x_\rho} G_{x\alpha}(ux)G_{\nu\alpha}(-ux)
- \frac{\partial}{\partial x_\nu} G_{x\alpha}(ux)G_{\rho\alpha}(-ux)\right]
\nonumber\\
&+&{} \int_0^1\!udu\, \Big[G_{x\alpha}(ux)G_{\nu\alpha}(-ux)
  - G_{\nu\alpha}(ux)G_{x\alpha}(-ux)\Big]\,,
\ea
which can be expressed in terms of three-gluon operators using the QCD equations
of motion as follows.

We begin with the expression in the second line of \re{A1}:
\ba{A2}
\lefteqn{\hspace*{-1cm}
  \frac{\partial}{\partial x_\rho} G_{\mu\alpha}(ux)G_{\nu\alpha}(-ux)
- \frac{\partial}{\partial x_\nu} G_{\nu\alpha}(ux)G_{\rho\alpha}(-ux) }
\nonumber\\&=&
G_{\mu\alpha}\left[u\derleft_\rho\!-\!u\derright_\rho
-i\int_{-u}^u \!\!vdv\,G_{x\rho}(vx)\right]G_{\nu\alpha}(-ux)
-(\rho\leftrightarrow\nu)\,.
\ea
Following \cite{BB89}, we introduce a derivative over the
total translation
\ba{trans}
 \partial_\rho G_{\mu\alpha}(ux)G_{\nu\alpha}(-ux) &\equiv&
 \frac{\partial}{\partial y_\rho}
\Big[G_{\mu\alpha}(ux+y)G_{\nu\alpha}(-ux+y)\Big]_{\textstyle {{y\to0}}}
\nonumber\\
&=&
G_{\mu\alpha}(ux)\left[\derleft_\rho\!+\!\derright_\rho
-i\int_{-u}^u \!\!dv\,G_{x\rho}(vx)\right]G_{\nu\alpha}(-ux)
\ea
so that
\ba{A4}
\lefteqn{
x_\rho x_\nu\left[
\frac{\partial}{\partial x_\rho} G_{\mu\alpha}(ux)G_{\nu\alpha}(-ux)-
u \partial_\rho G_{\mu\alpha}(ux)G_{\nu\alpha}(-ux)\right]
-(\rho\leftrightarrow\nu) =}
\nonumber\\&=&
-2u G_{x\alpha}(ux)\derright_\alpha G_{\nu x}(-ux) -
 i\int_{-u}^u \!\!\!(u-v)\,dv\, G_{x\alpha}(ux) G_{x\nu}(vx)G_{x\alpha}(-ux)
\nonumber\\
&=& {}\!2u\partial_\alpha G_{x\alpha}(ux)G_{x\nu}(-ux)
\!-\!2u G_{x\alpha}(ux)\!\derleft_\alpha
G_{x\nu}(-ux)
+2iu\!\!\int_{-u}^u \!\!\!\!\!dv\, G_{x\alpha}(ux) G_{x\alpha}(vx)G_{x\nu}(-ux)
\nonumber\\
&&{}- i\int_{-u}^u \!\!(u-v)dv\, G_{x\alpha}(ux)G_{x\nu}(vx)G_{x\alpha}(-ux)
\ea
where we used the Bianchi identity $\derright_\nu \!G_{\rho\alpha}-\derright_\rho
\!G_{\nu\alpha} =
\,\derright_\alpha\! G_{\nu\rho}$ to arrive at the expression in the second
line.
Neglecting total derivatives and terms $\sim D_\alpha G_{\alpha\nu}$ this gives
\ba{Apart1}
\lefteqn{
 \int_0^1\!udu\,x_\rho \left[
  \frac{\partial}{\partial x_\rho} G_{x\alpha}(ux)G_{\nu\alpha}(-ux)
- \frac{\partial}{\partial x_\nu} G_{x\alpha}(ux)G_{\rho\alpha}(-ux)\right]=}
\\
&=& i\!\!\int_0^1\!\!udu\!\!\int_{-u}^u \!\!\!\!dv\,
\Big[2u G_{x\alpha}(ux) G_{x\alpha}(vx)G_{x\nu}(-ux) +(v-u)
     G_{x\alpha}(ux)G_{x\nu}(vx)G_{x\alpha}(-ux)\Big].
\nonumber
\ea

Next, we consider the antisymmetric combination of gluon fields in the
last line in \re{A1} and rewrite it as
\ba{A5}
 \lefteqn{
 G_{\mu\alpha}(x)G_{\nu\alpha}(-x) - (\mu\leftrightarrow\nu)=}
\nonumber\\&=&
\int_0^1\!\!\frac{du}{u} x_\lambda\frac{\partial}{\partial x_\lambda}
\Big[G_{\mu\alpha}(ux)G_{\nu\alpha}(-ux) - (\mu\leftrightarrow\nu)\Big]
\nonumber\\
&=& \int_0^1\!\!du x_\lambda\Big[
 G_{\mu\alpha}(ux)\derleft_\lambda G_{\nu\alpha}(-ux)-
 G_{\mu\alpha}(ux)\derright_\lambda G_{\nu\alpha}(-ux)\Big]-
(\mu\leftrightarrow\nu)\,.
\ea
By a repeated application of the the Bianki identity and
separating (and then neglecting) contributions of total translation and
equation of motion terms $\sim D_\alpha G_{\alpha\nu}$, one arrives
after some algebra at the following expression:
\ba{A6}
 x_\mu\Big[G_{\mu\alpha}(x)G_{\nu\alpha}(-x) - (\mu\leftrightarrow\nu)\Big]
&=&
 2i\!\!\int_0^1\!\!du\!\!\int_{-u}^u \!\!\!\!dv\,
\Big[G_{x\alpha}(ux) G_{x\alpha}(vx)G_{x\nu}(-ux)
\nonumber\\&&{}\hspace*{-3cm}{}
     -G_{x\alpha}(ux) G_{x\nu}(vx)G_{x\alpha}(-ux) +
      G_{x\nu}(ux) G_{x\alpha}(vx)G_{x\alpha}(-ux)\Big]
\ea
that yields
\ba{Apart2}
\lefteqn{
\int_0^1\!udu\, \Big[G_{x\alpha}(ux)G_{\nu\alpha}(-ux)
  - G_{\nu\alpha}(ux)G_{x\alpha}(-ux)\Big]=}
\nonumber\\
&=&\int_0^1\!du\,(1-u^2)\!\!
\int_{-u}^u \!\!\!\!dv\,
\Big[G_{x\alpha}(ux) G_{x\alpha}(vx)G_{x\nu}(-ux) -
      G_{x\alpha}(ux) G_{x\nu}(vx)G_{x\alpha}(-ux) \mbox{\hspace*{1cm}}
\nonumber\\&&
\mbox{\hspace*{3.6cm}}+
      G_{x\nu}(ux) G_{x\alpha}(vx)G_{x\alpha}(-ux)\Big].
\ea
Taking the sum of the expressions in \re{Apart1}, \re{Apart2} and the
color trace, we obtain the result \re{trick3} quoted in the text.

\end{document}